\documentclass[11pt,a4paper]{article}
\pdfoutput=1
\usepackage{jcappub}

\usepackage{bm}
\usepackage{dsfont}
\usepackage{bbm}
\usepackage{color}

\allowdisplaybreaks[1]

\newcommand{\ii}{\mathbbm{i}}

\newcommand{\dd}{\text{d}}

\renewcommand\({\left(}
\renewcommand\){\right)}
\renewcommand\[{\left[}
\renewcommand\]{\right]}
\newcommand{\CLASS}{{\sc class}}
\newcommand{\CAMB}{{\sc camb}}
\newcommand{\CMBFAST}{{\sc cmbfast}}
\newcommand{\CMBEASY}{{\sc cmbeasy}}

\newcommand{\diff}[2]{\frac{\dd #1}{\dd #2}}

\newcommand{\kk}{\hat{K}}

\newcommand{\genbes}{\mathfrak{j}_\ell ^\nu}
\newcommand{\dgenbes}{\mathfrak{j}_\ell ^{\nu\prime}}
\newcommand{\ddgenbes}{\mathfrak{j}_\ell ^{\nu\prime\prime}}
\newcommand{\gencsck}{\mathfrak{csc}_K}
\newcommand{\gencotk}{\mathfrak{cot}_K}

\newcommand{\cotK}{\cot _K}
\newcommand{\sinK}{\sin _K}
\newcommand{\cscK}{\,\text{csc} _K}

\newcommand{\atan}{\,\text{atan}}

\newcommand{\asinh}{\,\text{asinh}}
\newcommand{\asin}{\,\text{asin}}

\begin{document}
\hypersetup{pageanchor=false}
\hfill CERN-PH-TH/2013-298, LAPTH-071/13

\title{Fast and accurate CMB computations in non-flat FLRW universes}
\author[a,b,c]{Julien Lesgourgues}
\author[a]{and Thomas Tram}
\affiliation[a]{Institut de
Th\'eorie des Ph\'enom\`enes Physiques, \'Ecole Polytechnique
F\'ed\'erale de Lausanne, CH-1015, Lausanne,
Switzerland}
\affiliation[b]{CERN, Theory Division, CH-1211 Geneva 23, Switzerland}
\affiliation[c]{LAPTh (CNRS - Universit\'e de Savoie), BP 110, F-74941 Annecy-le-Vieux Cedex, France}
\emailAdd{thomas.tram@epfl.ch}
\emailAdd{Julien.Lesgourgues@cern.ch}

\abstract{We present a new method for calculating CMB anisotropies in a non-flat Friedmann universe, relying on a very stable algorithm for the calculation of hyperspherical Bessel functions, that can be pushed to arbitrary precision levels.  We also introduce a new approximation scheme which gradually takes over in the flat space limit and leads to significant reductions of the computation time.

Our method is implemented in the Boltzmann code \CLASS{}. It can be used to benchmark the accuracy of the \CAMB{} code in curved space, which is found to match expectations. For default precision settings, corresponding to 0.1\% for scalar temperature spectra and 0.2\% for scalar polarisation spectra, our code is two to three times faster, depending on curvature. We also simplify the temperature and polarisation source terms significantly, so the different contributions to the $C_\ell $'s are easy to identify inside the code. 
}
\maketitle

\hypersetup{pageanchor=true}
\section{Introduction}

The large amount of cosmological information stored in the statistical properties of Cosmic Microwave Background (CMB) anisotropies and of the large scale structure of the universe has motivated several ambitious space-based and ground-based experiments. The interpretation of the data relies on a comparison with theoretical predictions, computed in the linear regime by sophisticated Bolztmann codes. In order to trust this interpretation, one should keep asking two questions: (i) are we postulating the correct model to describe the evolution of the universe? (ii) for a given model, are theoretical predictions calculated with sufficient accuracy, given observational errors? While the first question can be addressed by a continuous theoretical effort and by evaluating the goodness of fit of different models given the data, the second question calls for several independent tests of the robustness of Boltzmann codes. On top of that, progress in these codes are triggered by the fact that their speed is crucial, given that cosmological parameter extraction relies on Monte Carlo methods. These require the evaluation of tens of thousands of models each time that one cosmology is compared to one data set. In the global analysis of a data set like Planck \cite{Ade:2013ktc}, many cosmologies and combinations of data need to be considered, leading to millions of Boltzmann code runs.

\subsection{Current codes and data}
Several Boltzmann codes have been made public and compared with each other, including \CMBFAST{}, \cite{Seljak:1996is,Zaldarriaga:1996xe,Zaldarriaga:1997va}, implementing for the first time the line-of-sight method, and later \CAMB{}~\cite{Lewis:1999bs}, \CMBEASY{}~\cite{Doran:2003sy} and \CLASS{} \cite{Lesgourgues:2011re,Blas:2011rf}. Both \CAMB{} and \CLASS{} are being maintained and pushed to ever higher precision, and both were used in different parts of the Planck data analysis. 

\begin{figure}\label{fig:decomposition}%
\includegraphics[width=0.5\columnwidth]{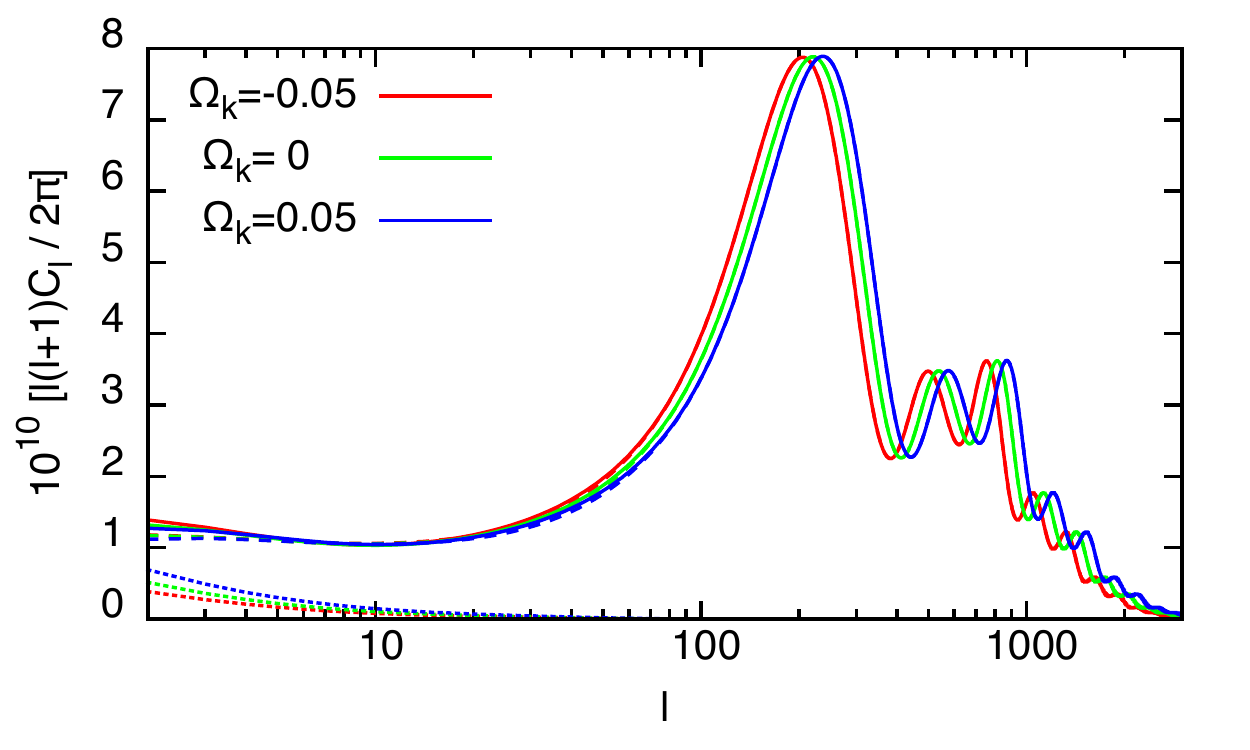}%
\includegraphics[width=0.5\columnwidth]{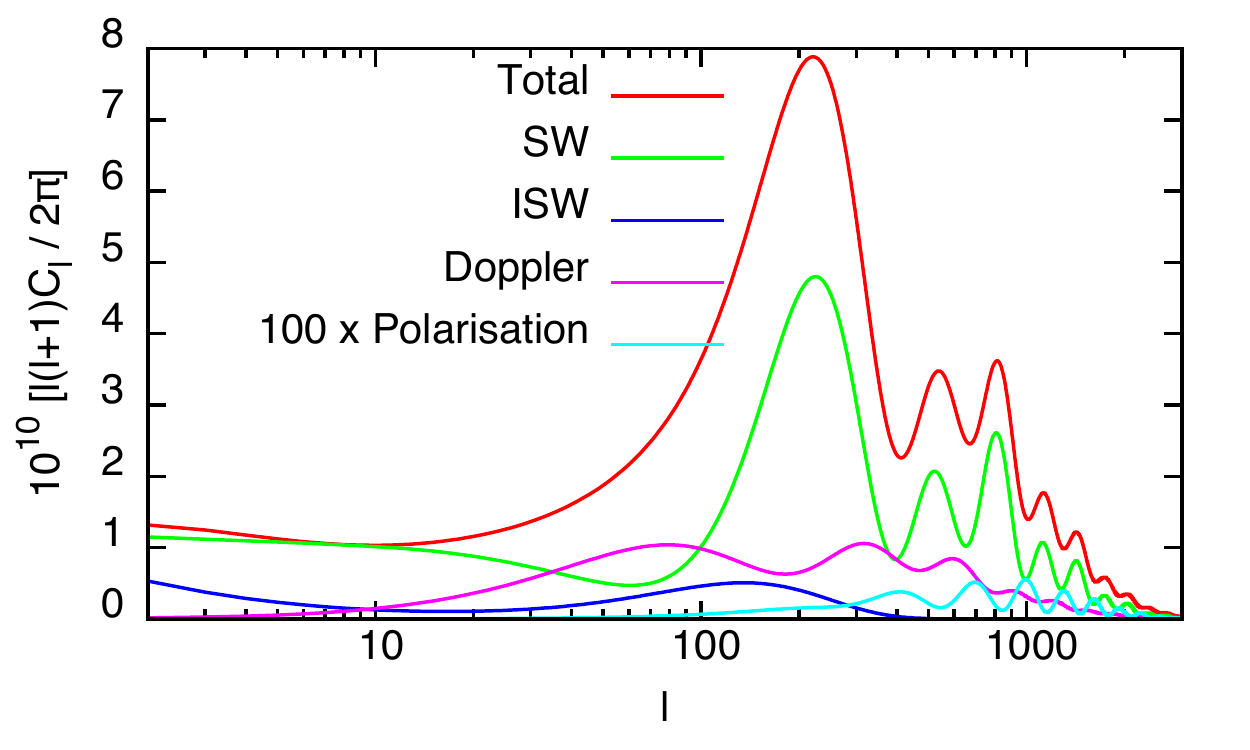}%
\caption{{\it(Left panel)} Temperature anisotropy spectrum for some closed, flat and open models ($\Omega_k=-0.05, 0, 0.05$, with all other cosmological parameters kept fixed). In each case, we show the auto-correlation spectrum of the total temperature anisotropy {\it (solid)}, of the late Integrated Sachs-Wolfe (ISW) contribution alone {\it (dotted)}, and of the total minus this contribution {\it (dashed)}. {\it(Right panel)} For a flat model, auto-correlation spectrum of the total temperature anisotropy and of different contributions (intrinsic temperature corrected by the Sachs-Wolfe term (SW), early plus late ISW term, Doppler term, polarisation contribution enhanced by a factor $10^2$). Such a decomposition can be obtained using the latest \CLASS{} version without any modifications (see Appendix A). Note that the individual auto-correlation spectra do not sum up to the total one, due to the existence of cross-correlation spectra.}%
\end{figure}

All current cosmological data can be well fitted with a flat FLRW model. Still, it is important to have efficient Boltzmann codes which also cover non-flat models, in order to check for small deviations from spatial flatness in future data. Currently, the bounds read $100\Omega_k=-1^{+1.8}_{-1.9}$ at the 95\% confidence level using Planck alone and $100\Omega_k=-0.10^{+0.62}_{-0.65}$ when combining Planck with BAO~\cite{Ade:2013zuv}. The effect of curvature on the CMB consists mainly in a shift of the angular scale of the acoustic peaks, due to a modification of the angular diameter distance to recombination. It also impacts the primordial spectrum and the evolution of very large wavelengths comparable to the curvature radius of the universe, especially through the late integrated Sachs-Wolfe effect (see Figure~\ref{fig:decomposition}, left panel).

The computation of temperature and polarisation power spectra in non-flat FLRW universes requires the calculation of hyperspherical Bessel functions when employing the line-of-sight method. Because of memory restrictions, these functions must be computed on the fly, and this imposes an execution penalty when analysing non-flat models. Spatial curvature was successfully implemented in both \CMBFAST{} and \CAMB{}, although both codes run significantly slower than in flat models. Their accuracy is also not well tested, due to the lack of a robust and accurate way of calculating hyperspherical Bessel functions without making any approximation. 

To this end, we implement a new and improved method for computing the hyperspherical Bessel functions which is fast and accurate. 
We also found a new approximation scheme which gradually takes over in the flat limit $K\rightarrow 0$. This scheme results in a significant speed-up for nearly flat models, and it ensures continuity in the point $K=0$ when doing parameter estimation. Our approach is implemented in the release 2.0 of the \CLASS{} code\footnote{To be precise, this paper always refer to the version with release number 2.0.4. The first \CLASS{} release including spatial curvature was 2.0.0. It was quickly followed by some minor revisions fixing small bugs, and at the time of submitting this work, by the revision 2.0.4, in which we improved the sampling scheme in wavenumber space and the tuning of some accuracy parameters, corresponding exactly to the results presented in this paper.}.

\subsection{Outline of the paper}
In section~2, we describe our approach for calculating hyperspherical Bessel functions. We also show a comparison with other methods, in order to estimate its accuracy on robust ground. In section~3, we describe other aspects of our implementation of curvature in a Boltzmann code. We explain how and why we simplified the temperature and polarisation source terms. We write the photon transfer functions, featuring source functions and radial functions, in a way which unifies flat and non-flat calculations. We clarify the issue of how to define primordial spectra and to integrate  over wavenumbers in non-flat space. In section~4, we describe our new approximation scheme for speeding up calculations in curved space, the flat rescaling approximation. In section~5, we evaluate the accuracy of our implementation for ``reference precision'' and ``default precision'' setting. We perform a detailed comparison of our results with those of \CAMB{} (version of November 2013) and compare the performances of the two codes. We present our conclusions in section~6. In appendix A, we show how to use \CLASS{} to split the temperature power spectrum into different contributions as in figure~\ref{fig:decomposition}. Finally, in appendix B and C, we summarise two other approximation schemes which were also introduced in the new version of \CLASS.

\section{Hyperspherical Bessel functions\label{sec:hyper}}
Here we give a brief overview of how we compute the hyperspherical Bessel functions in \CLASS{} \cite{Blas:2011rf,Lesgourgues:2011re}. The algorithms are given in much more detail in the paper~\cite{Tram:2013xpa}.
\subsection{Definition and notation}
In order for us to give unified formulae for any $K$, it is convenient to introduce the notation
\begin{align}
\sinK(\chi) &\equiv \frac{1}{\cscK \chi} \equiv \left\{
\begin{array}{ll}
\sinh \chi    & \kk=-1 \\
\chi          & \kk=0\\
\sin \chi     & \kk=1
\end{array}
\right. , &
\cotK (\chi) &\equiv \left \{ 
\begin{array}{ll}
\coth \chi    & \kk=-1 \\
\frac{1}{\chi}& \kk=0\\
\cot \chi     & \kk=1
\end{array}
\right. .
\end{align}
We use $\kk \equiv K/|K|$ for $K\neq 0$, otherwise $\hat{K}=0$. The three cases $\kk=0,1,-1$ refer respectively to a universe with null, elliptic or hyperbolic curvature, or in more common words, to a flat, closed or open Universe. The hyperspherical Bessel functions $\Phi^\nu_\ell (\chi)$ are given by $\Phi^\nu_\ell (\chi) = \cscK(\chi) u^\nu_\ell (\chi)$, where $u^\nu_\ell (\chi)$ is the solution of
\begin{equation}
\diff{^2 u^\nu_\ell }{\chi^2} = \[ \ell (\ell +1)\cscK^2(\chi) - \nu^2\] u^\nu_\ell (\chi),
\label{eq:u_equation}
\end{equation}
which is regular at the origin. For $\kk=0$ the hyperspherical Bessel functions are just the usual spherical Bessel functions, $\Phi^\nu_\ell (\chi) = j_\ell (\nu \chi)$, while for $\kk \pm 1$, they can be expressed in terms of Legendre functions~\cite{Harrison:1967,Tram:2013xpa}:
\begin{equation}\label{eq:phiLegendre}
\Phi^\nu_\ell (\chi) = \left\{
\begin{array}{ll}
\(\prod\limits_{n=1}^\ell {\sqrt{\nu^2+n^2}} \) \sqrt{\frac{\pi }{2\sinh \chi}} P^{-1/2-\ell }_{-1/2+\ii\nu}(\cosh\chi) 	&\kk=-1\\
\( \prod\limits_{n=1}^\ell {\sqrt{\nu^2-n^2}} \) \sqrt{\frac{\pi}{2\sin \chi}}  P^{-1/2-\ell }_{-1/2+\nu}(\cos \chi) 	&\kk=1\\
\end{array}
\right. .
\end{equation}
When $\kk=1$ we must also require the solution to be regular at the second boundary $\chi=\pi$. This leads to the requirement that $\nu$ must be an integer, so the eigenmode spectrum becomes discrete~\cite{Tram:2013xpa} which is the familiar energy quantisation for a particle trapped in a potential. It also has the consequence that the solutions become symmetric (anti-symmetric) around $\chi=\frac{\pi}{2}$ for $\nu-\ell -1$ even (odd), so we only need to compute the solution in the range $[0; \pi/2]$. 
However we found that this has one more important consequence, namely, that
%The most important consequence however, is that 
the $\kk=1$ hyperspherical Bessel functions can be re-expressed in terms of Gegenbauer polynomials $C_\alpha^{(\beta)}(x)$~\cite{Tram:2013xpa}:
\begin{equation}\label{eq:Gegenbauer_identity}
\Phi^\nu_\ell (\chi) = 2^\ell  \ell ! \sqrt{\frac{(\nu-\ell -1)!}{\nu(\nu+\ell )!}} \sin^\ell (\chi) C_{\nu-\ell -1}^{(\ell +1)}(\cos \chi).
\end{equation}
This identity can be used directly in a compact, stable method for $\kk=1$ hyperspherical Bessel functions, or it can be used for solving the long-standing issue~\cite{Kosowsky:1998nc} of using backward recurrence for $\kk=1$.

\subsection{Establishing a reference method}
\begin{figure}\label{fig:reldif}%
\includegraphics[width=0.5\columnwidth]{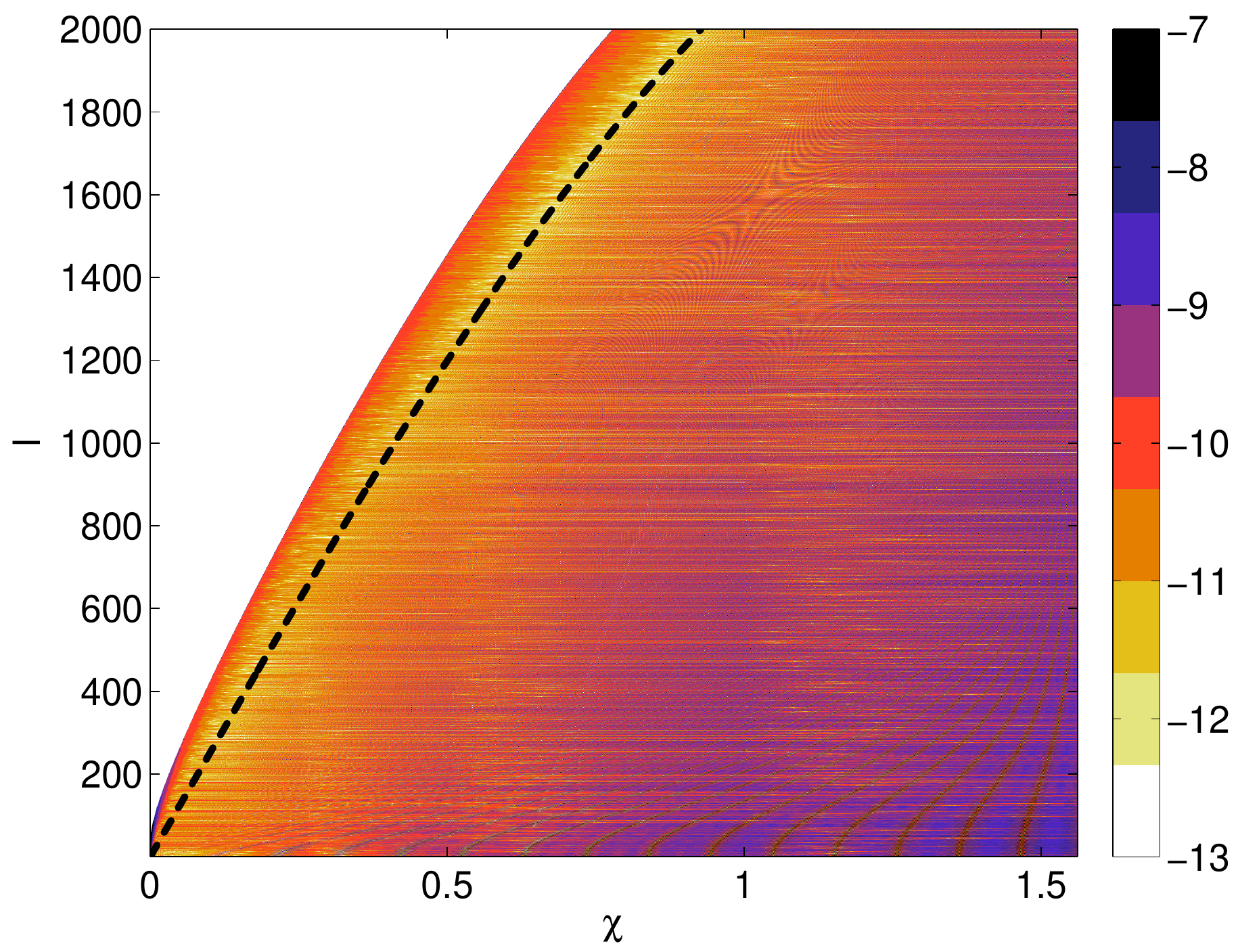}%
\includegraphics[width=0.5\columnwidth]{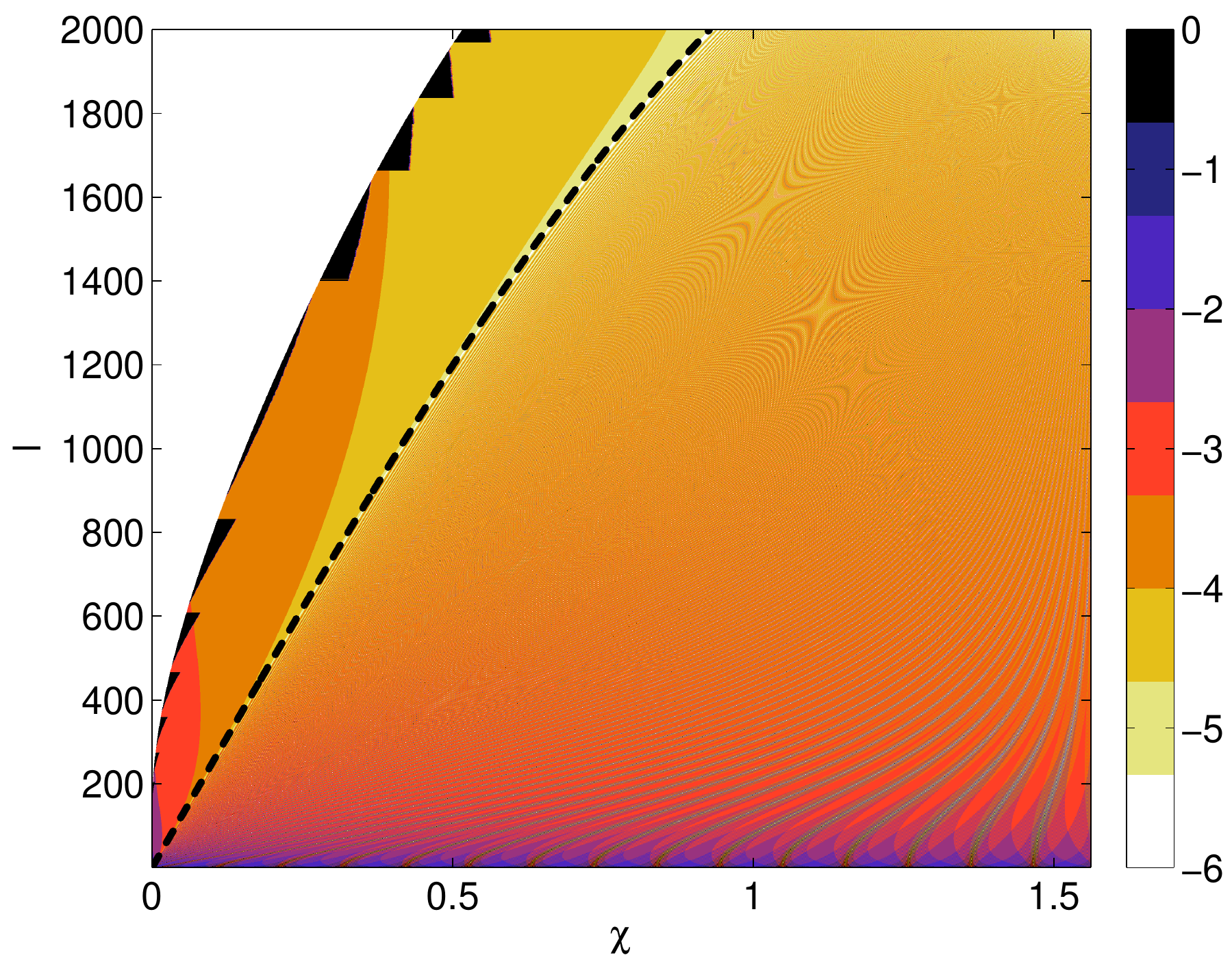}%
\caption{Logarithm  (base 10) of the relative difference between different computations of the $K=1$, $\nu=2500$ hyperspherical Bessel functions. The dashed line denotes the position of the turning point. \emph{Left panel:} Direct integration of the differential equation using the modified Magnus method compared to the Gegenbauer polynomial method. \emph{Right panel:} Relative error of the WKB approximation method compared to the Gegenbauer polynomial method.}%
\end{figure}

The first step in building the hyperspherical Bessel module was to find a reliable reference implementation to which we could compare our methods. From equation~\eqref{eq:phiLegendre} we see that they are given in terms of Legendre functions and also hypergeometric functions using well-known identites. But no general and stable numerical algorithm exist for the computation of Legendre functions or hypergeometric functions of large order, not even in commercial software. So for establishing a set of trusted methods, we implemented many different methods in MATLAB. Whenever possible we coded the routines such that they would work also for $\kk=0$: the first check would then be to test the routine against MATLAB's built-in Bessel function routine which is known to be precise. We implemented the following 5 schemes in MATLAB:
\begin{enumerate}
	\item {\bf Direct evaluation of equation~\eqref{eq:phiLegendre}.} The Legendre functions in equation~\eqref{eq:phiLegendre} can be rewritten as hypergeometric functions and evaluated using the hypergeometric series. While initially promising, the series will not converge numerically for large $\ell $ since the positive and negative part of the series become large and nearly equal.
	\item {\bf Direct integration of the differential equation~\eqref{eq:u_equation}.} Such an implementation starts with an arbitrary, finite initial condition deep inside the dissipative region where the solution is heavily damped. It then evolves into the dispersive regime where the solution oscillates, and the solution will be the regular one up to an overall normalisation factor. This factor is then typically fixed by a forward recurrence. The standard ODE-methods works well for low to medium accuracy, but for high accuracy we implemented the modified Magnus method~\cite{Iserles:2002:GED}.
	\item {\bf WKB approximation.} Equation~\eqref{eq:u_equation} lends itself to a WKB approximation~\cite{Kosowsky:1998nc,Tram:2013xpa}. While this approximation is not sufficiently accurate as a \CLASS{} reference method, it is an important cross check. We also expected this method to be faster than the recurrence method for standard precision, but that turned out not to be the case\footnote{Each evaluation of the WKB approximation requires two calls to library functions and a Chebyshev approximation of the Airy function. While it is faster than the recurrence method for a single $\ell $-value, it becomes slower when we need a few hundred $\ell 's$ for each $\nu$ which is usually the case.}.
	\item {\bf Using recurrence relations.} Forward recurrence is only stable inside the dispersive region, so in the dissipative region one must use backward recurrence instead. It was thought for a long time~\cite{Kosowsky:1998nc} that backward recurrence could not be used for $\kk=1$ due to the restriction $\ell <\nu$, but this problem is eliminated by the identity we found, equation~\eqref{eq:Gegenbauer_identity}.
	\item {\bf Using the Gegenbauer identity directly.} Equation~\eqref{eq:Gegenbauer_identity} gives the $\kk=1$ solutions in terms of Gegenbauer polynomials which we compute using (stable) forward recurrence.\footnote{Note that this is different from using the Gegenbauer identity to set initial conditions for the backward recurrence, since the Gegenbauer recurrences and hyperspherical recurrences moves along different lines in the $(\ell,\nu)$-plane.}
\end{enumerate}
We compared all the methods on the parameter space consisting of $\ell $, $\nu$ and $\chi$. For instance, we selected $N$ $\chi$-samples in a prescribed way and generated a figure of $N$ subplots, each containing a 2D image of the logarithm of the relative error between two methods. Similarly, we kept $\nu$ fixed and plotted the logarithm of the relative difference between methods in the $(x,\ell )$-plane, and we show such a plot for $\kk=1$ for a fixed $\nu=2500$ in figure~\ref{fig:reldif}. The left panel compares method 2 and 5, while the right panel compares the WKB method 3 to method 5. This is probably the first time that the WKB approximation to the hyperspherical Bessel function has been properly quantified, and we can see that the error is usually of the order $10^{-4}$. The dashed black line shows the position of the classical turning point, so the solutions decay exponentially to the left of this line.

The method implemented in \CAMB{} is the following: the differential equation is integrated from an initial condition computed primarily using the WKB approximation and sometimes from the recurrence relations. The integration method used is the fourth order explicit Runge-Kutta method, RK4. Unlike the usual embedded methods, RK4 has no built-in error estimate of the local truncation error for adaptive step size, so the step sizes are taken from the time sampling of sources but with a hard coded maximum step. The differential equation is solved in tandem with the source convolution which makes it difficult to judge the accuracy of the actual spherical Bessel functions. It is also clear that this method will usually be limited by the accuracy limit of the WKB approximation.

Some but not all of the hyperspherical Bessel implementation in \CAMB{} is inherited from \CMBFAST{}~\cite{Zaldarriaga:1997va}. This includes the idea of using RK4 and summing up the source convolutions simultaneously. However, \CMBFAST{} found initial values from interpolation in a precomputed table, while \CAMB{} is using recurrence and WKB.

\subsection{Choosing a scheme for CLASS}\label{sec:CLASS_scheme}
The differential equation method has the advantage that one gets all $\chi$-values with a single call to the method, while the recurrence method will generate all $\ell $-values with a single call. Both methods will generate the derivative ${\Phi^\nu_\ell }^\prime(\chi)$ for free, which will be important later for interpolation. One can argue for both methods, but we chose to implement the recurrence method because it is more standard, and it would be needed anyway for normalising the ODE-method. It also has the advantage of requiring a negligible amount of library function calls such as trigonometric functions and square roots, so it would possibly be easier to vectorise.

\begin{figure}[htb]
\begin{center}
\includegraphics[width=0.8\columnwidth]{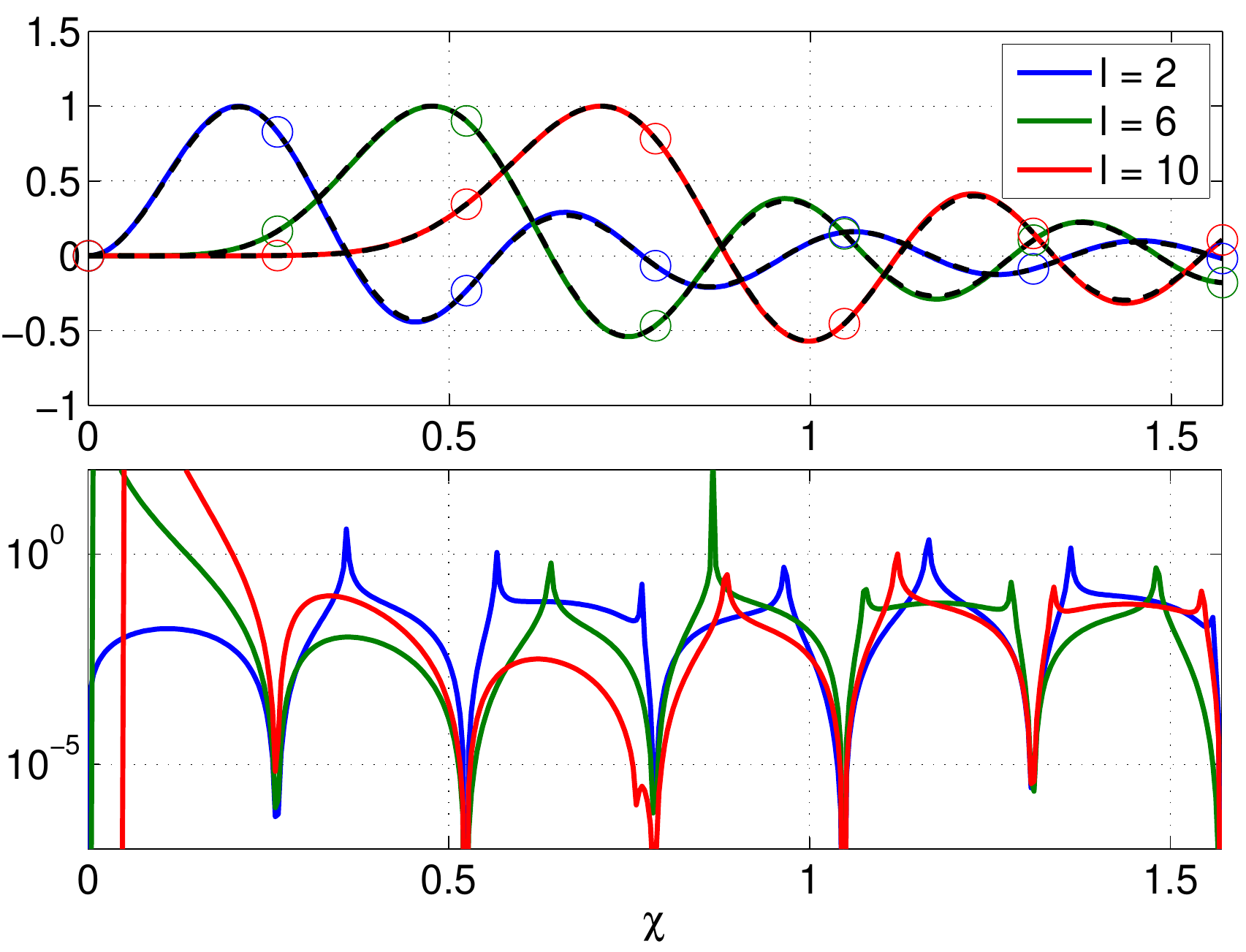}%
\end{center}
\caption{Hermite interpolation of $\kk=-1$, $\nu=16$ hyperspherical Bessel functions for 3 values of $\ell $. Only $7$ points are used, corresponding to a sampling parameter of $1.75$. The top plot shows the exact functions in solid lines and the interpolated functions in black dashed lines. The bottom plot shows the relative error of the interpolated function, which is of the order $10^{-2}$ despite the very low value of the sampling parameter.}
\label{fig:interpolation}%
\end{figure}

We found the recurrence method to be accurate at the level of $\sim 10^{-10}$, which is of course much better than what is required for standard precision in \CLASS{}. To extend the dynamic range of the method (i.e. increasing speed by reducing accuracy), we coded an interpolation method based on Hermite interpolation. (See e.g.~\cite{burden:2004} and the help entry for the Mathematica function \texttt{InterpolatingPolynomial}.) By storing just $\Phi^\nu_\ell $ and ${\Phi^\nu_\ell }^\prime$, all higher derivatives can be found from the differential equation and its derivatives. We then construct the unique order 5 polynomial which matches $\Phi^\nu_\ell $, ${\Phi^\nu_\ell }^\prime$ and ${\Phi^\nu_\ell }^{\prime\prime}$ inside each interval for interpolating $\Phi^\nu_\ell $. However, since we sometimes also need to interpolate ${\Phi^\nu_\ell }^\prime$ and ${\Phi^\nu_\ell }^{\prime\prime}$, we need formulae for ${\Phi^\nu_\ell }^{\prime\prime\prime}$ and ${\Phi^\nu_\ell }^{\prime\prime\prime\prime}$ as well. We refer the interested reader to~\cite{Tram:2013xpa} for more details on the implementation.

The interpolation method is illustrated in figure~\ref{fig:interpolation} for a sampling parameter of $1.75$. This parameter is defined as the number of points per approximate wavelength $\frac{2\pi}{\nu}$, so with 1.75 we compute less than two points per oscillation. Still the relative error for such a sampling parameter is as low as $\sim 10^{-2}$. Increasing the sampling gradually reduces the relative error all the way to a sampling of $\sim 30$ where the relative error due to the interpolation becomes $10^{-10}$, i.e. comparable to the accuracy of the underlying method. A typical sampling value of $5.0$ yields a relative error of $\sim 10^{-5}$. We will see at the end of section~\ref{sec:transfer} how this interpolation scheme is implemented in \CLASS{}.

\section{Efficient CMB spectrum calculations in curved space}

In this paper and inside the code, we follow the notations of Ma \& Berstchinger~\cite{Ma:1995ey} for metric perturbations, matter perturbations and Bolztmann hierarchies (see~\cite{Tram:2013ima} for the generalisation of these hierarchies to curved space, and for vector/tensor mode notations). In this paper, we write equations only in the Newtonian gauge, although in \CLASS{} they are implemented in both the Newtonian and synchronous (comoving with CDM) gauges. It is trivial to infer the synchronous gauge formulas using the gauge transformation rules presented in~\cite{Ma:1995ey}.

\subsection{Source functions\label{sec:sources}}
It is well-known since the work of~\cite{Seljak:1996is} that photon transfer functions can be conveniently obtained by a time integral of a set of source functions multiplied by certain radial functions. The total angular momentum method introduced in~\cite{Hu:1997hp} (and generalised to curved space in~\cite{Hu:1997mn}) leads to a unified set of simple integrals, accounting for temperature $T$ (decomposed in three contributions $T_j$ with $j=0,1,2$), E-type polarisation $E$, and B-type polarisation $B$: 
\begin{eqnarray}
{\Delta_\ell^{T_j}}^{(m)}(q) &=& \int_{\tau_\mathrm{ini}}^{\tau_0} d\tau S_{T_j}^{(m)}(k,\tau) \phi^{jm}_{\ell}(\nu,\chi)~, \label{eq:transT}\\
{\Delta_\ell^{E}}^{(m)}(q) &=& \int_{\tau_\mathrm{ini}}^{\tau_0} d\tau S_{P}^{(m)}(k,\tau) \epsilon^{m}_{\ell}(\nu,\chi)~, \\
{\Delta_\ell^{B}}^{(m)}(q) &=& \int_{\tau_\mathrm{ini}}^{\tau_0} d\tau S_{P}^{(m)}(k,\tau) \beta^{m}_{\ell}(\nu,\chi)~. \label{eq:transB}
\end{eqnarray}
The $\Delta$'s are the photon transfer functions, depending on the mode index $m$ ($m=0,1,2$ for scalars, vectors, tensors), on the multipole $\ell$ and on the generalised wavenumber 
\begin{equation}
q=\sqrt{k^2 + (1+m)K}~, \qquad K=-H_0^2(1-\Omega_\mathrm{tot})~.
\end{equation}
The radial functions $\{ \phi^{jm}_{\ell}, \epsilon^{m}_{\ell}, \beta^{m}_{\ell} \}$ are linear combinations of (hyper-)spherical Bessel functions and their derivatives. They depend on the rescaled generalised wavenumber $\nu=q/\sqrt{|K|}$ and the rescaled radial coordinate $\chi\equiv \sqrt{|K|} (\tau_0-\tau)$. This remains true in the flat limit as well, since the hyperspherical Bessel functions becomes $\Phi^\nu_\ell (\chi) \rightarrow j_\ell (\nu \chi) = j_\ell (k (\tau_0-\tau)$ so the two $\sqrt{|K|}$'s cancels out. All the relevant radial functions in flat space may be found in~\cite{Hu:1997hp}, while the $\kk=-1$ radial functions can be found in~\cite{Hu:1997mn}. In section~\ref{sec:radial} we write all the radial functions in a unified form where the flat limit is transparent.

The $S$'s are the source functions which depends on the Fourier wavenumber $k(q)$ and on conformal time $\tau$. In the Newtonian gauge, and omitting vector modes, the source functions derived in~\cite{Hu:1997hp} read
%This is bad LaTeX for so many reasons!
%\begin{eqnarray}
%S_{T0}^{(0)}&=&\frac{g}{4}\delta_\gamma^{(0)} + e^{-\kappa} \phi'~,~~~~
%S_{T1}^{(0)}= \frac{g}{k} \theta_b^{(0)} + e^{-\kappa} k \psi ~,~~~~
%S_{T2}^{(0)}=g P^{(0)}~, \label{eq:sourceT}\\
%S_P^{(0)} &=& \sqrt{6} g P^{(0)}  ~,~~~~~~~~~
%S_{T2}^{(2)} = g P^{(2)} -e^{-\kappa}h' ~,~~~
%S_P^{(2)} = \sqrt{6} g P^{(2)}~,
%\end{eqnarray}
\begin{align}
S_{T0}^{(0)} &=\frac{g}{4}\delta_\gamma^{(0)} + e^{-\kappa} \phi'~, &
S_{T1}^{(0)} &= \frac{g}{k} \theta_b^{(0)} + e^{-\kappa} k \psi ~, &
S_{T2}^{(0)} &= g P^{(0)}~, \label{eq:sourceT}\\
S_P^{(0)}    &= \sqrt{6} g P^{(0)}  ~, &
S_{T2}^{(2)} &= g P^{(2)} -e^{-\kappa}h' ~, &
S_P^{(2)}    &= \sqrt{6} g P^{(2)}~,
\end{align}
while $S_{T0}^{(2)}=S_{T1}^{(2)}=0$.
The polarisation source terms $P^{(0)}$ and $P^{(2)}$ can be expressed in terms of the coefficients of the optimal Boltzmann hierarchies presented in~\cite{Tram:2013ima}. They are given in equation~(2.16) of that reference but we will reprint them here for convenience:
\begin{align}
P^{(0)} &= \frac{1}{8}\left[F_{\gamma 2}^{(0)}+G_{\gamma 0}^{(0)}+G_{\gamma 2}^{(0)}\right]~,\\
P^{(2)} &= -\frac{1}{\sqrt{6}} \left[ \frac{1}{10} F_{\gamma 0}^{(2)} + \frac{1}{7} F_{\gamma 2}^{(2)} + \frac{3}{70} F_4^{(2)}
- \frac{3}{5} G_{\gamma 0}^{(2)} + \frac{6}{7} G_{\gamma 2}^{(2)} - \frac{3}{70} G_{\gamma 4}^{(2)}\right]~.
\end{align}
Note that we extend the naming scheme of~\cite{Ma:1995ey} for the lowest multipoles such that $F_{\gamma 2}^{(0)} = 2 s_2 \sigma_{\gamma 2}^{(2)}$, $F_{\gamma 2}^{(2)} = 2 \sigma_\gamma^{(2)}$ and $F_{\gamma 0}^{(m)} = \delta_\gamma^{(m)}$.

In previous codes (\CMBFAST{}~\cite{Seljak:1996is}, \CAMB{}~\cite{Lewis:1999bs}, \CMBEASY{}~\cite{Doran:2003sy} and \CLASS{} up to version 1.7), the temperature source functions were not implemented as such. They were integrated by part (once for the $T_1$ and twice for $T_2$), in order to reduce the problem to the integration of a single source $S_T$ multiplied by a single Bessel function. This approach is not very practical, because terms like e.g. $\psi'$ (or its counterpart in the synchronous gauge\footnote{In the synchronous gauge, $\psi$ corresponds to $\alpha'+\frac{a'}{a} \alpha$, with $\alpha \equiv (h'+6\eta')/(2k^2)$. Hence $\psi'$ involves $\alpha''$.}) are not directly given by Einstein equations. One needs to evaluate the time derivative of these equations, which involves the time derivative of the pressure $\bar{p}_i$ and anisotropic stress $\sigma_i$
of photons and baryons. This can be done in two ways:
\begin{itemize}
\item in \CMBFAST{}, \CAMB{} and \CMBEASY{}, all these derivatives are calculated exactly using the equations of motion.
As a result, the evaluation of $S_T$ requires several extra equations (e.g. for computing the derivative of massive neutrino pressure). This causes the final expression of the temperature source function to be very lengthy and difficult to understand. For instance, in this scheme it is non-trivial to split the source term into the different contributions: Sachs-Wolfe (SW), Integrated Sachs-Wolfe (ISW), Doppler, etc.
\item in \CLASS{} up to version 1.7, the code computed $S_{T_0}$, $S_{T_1}^\prime$ and $S_{T_2}^\prime$ exactly, while $S_{T_2}^{\prime \prime}$ was inferred from $S_{T_2}^\prime$ using a numerical derivative. This scheme enabled relatively simple and compact formulas without derivatives of Einstein equations, but some fine-tuning of precision parameters was required for computing the derivative with sufficient accuracy.
\end{itemize}
In \CLASS{} version 2.0 and higher, the strategy consists in keeping all three terms $S_{T_0}$, $S_{T_1}$, $S_{T_2}$ for the scalar temperature source functions. They lead to the three transfer functions ${\Delta_{\ell}^{T_j}}^{(m)}(q)$ introduced in equation~(\ref{eq:transT}), which are summed up in the module called ``spectra'', to obtain a single temperature transfer function ${\Delta^T_{\ell}}^{(m)}(q)$. The advantage of this method is that we maintain high-accuracy while sticking to compact and simple source functions. The drawback is that the number of transfer functions increases from 1 to 3 when CMB temperature is required, but this actually has a minor impact on the performance of the code since 2 of the 3 contributions can be computed quickly using the approximations in appendix B and C.

We can still perform some integrations by part if we want to distribute the various terms differently between $S_{T_0}$, $S_{T_1}$, $S_{T_2}$. A good policy consists in using this freedom for optimising the speed and numerical stability of the code, up to the extent to which the formulas remain simple, and do not require derivatives of Einstein equations. This can be achieved by reorganising the terms in such a way that $S_{T_0}$, $S_{T_1}$, $S_{T_2}$ all become very small between recombination and reionisation. If this is the case, over this interval of time, we will not need a very accurate time-sampling of Bessel functions. This does not happen when one sticks to the original expressions (\ref{eq:sourceT}), in which $S_{T0}^{(0)}$ and $S_{T_1}^{(0)}$ both remain large after recombination, but nearly cancelling each other. To avoid this, we can introduce a new set of source functions: 
\begin{equation}
\tilde{S}_{T_0}^{(0)}=g\left( \frac{1}{4}\delta_\gamma+\phi\right)+e^{-\kappa} 2 \phi' + k^{-2}(g \theta_b'+g' \theta_b)~,~~~~
\tilde{S}_{T_1}^{(0)}=e^{-\kappa} k (\psi-\phi)~,~~~~
\tilde{S}_{T_2}^{(0)}=S_{T2}^{(0)}~.
\label{eq:sourceTbis}
\end{equation}
Using integrations by part, one can easily show that this choice is exactly equivalent to (\ref{eq:sourceT}). However $\tilde{S}_{T_1}^{(0)}$ is proportional to $(\phi-\psi)$, which is very small at any time inside the Hubble radius, due to the decay of all anisotropic stresses. Similarly, $\tilde{S}_{T_0}^{(0)}$ remains very small between reionisation and recombination because $g$ and $g'$ are small, and $\phi$ is nearly constant. This version of the source functions is both compact and numerically efficient, and is the one implemented in \CLASS\footnote{in the function {\tt perturb\_sources()} of the module {\tt perturbations.c}.}. Its counterpart in the synchronous gauge can be read directly in the code.  In appendix~\ref{app:sources}, we show how to break this expression into SW, ISW, Doppler and polarisation contributions.

\subsection{Computing the transfer functions \label{sec:transfer}}

In order to compute the functions $\phi_\ell^{j(m)}(\nu, \chi)$, $\epsilon_\ell^{(m)}(\nu, \chi)$ and $\beta_\ell^{(m)}(\nu, \chi)$, we need to know the hyperspherical Bessel functions $\Phi_\ell^\nu(\chi)$ and their first and second derivatives for several values of $\ell$, $\chi$ and $\nu$. For a given rescaled wavenumber $\nu$, a single run of the recurrence method summarised in section~\ref{sec:hyper} gives $\Phi_\ell^\nu$ and $\Phi_\ell^{\nu \prime}$ for all $\ell$, and the interpolation routine will compute the higher derivatives as they become necessary. In the ``transfer'' module of the code, i.e. the module in charge of performing the integrals of equations~(\ref{eq:transT}-\ref{eq:transB}), we wish to avoid any redundancy in the calculation of hyperspherical Bessel functions. Since we can not store them in memory, we are left with only one possible strategy:
\begin{itemize}
\item implement the loop over $q$ values (or equivalently over $\nu$ values) as the outermost loop in the ``transfer'' module.
\item for each new value of $q$ and $\nu$,  compute all hyperspherical Bessel functions for a discrete set of arguments $\{ \chi_i\}$, and store the result in a form that will allow fast interpolation at any arbitrary $\chi$. 
\item inside the $q$-loop, loop over modes (scalar, vectors, tensors), over initial conditions (adiabatic, isocurvature), over types of transfer functions (temperature $T_0$, $T_1$, $T_2$, polarisation $E$, $B$, and other types related to weak lensing and density  power spectra), and finally over multipoles $\ell $.
\end{itemize}
Since the calculation of Bessel functions is the most time-consuming part of the code for large curvature, it is crucial to run the recurrence for as few values of $\{ \chi_i\}$ as possible. This is where the interpolation scheme of section~\ref{sec:CLASS_scheme} comes into play. As explained in that section, we checked explicitly that by increasing the sampling parameter, we can obtain at least $10^{-10}$ precision on the hypersperical Bessel function, i.e. much more than will ever be needed by Boltzmann codes. We will show in section~\ref{sec:accuracy} how to obtain ``reference'' and ``default'' precision parameter settings, and which accuracy they imply for the $C_\ell$'s. We found that a sufficient choice for the sampling parameter in the default case is $6$ for $\nu<1000$, and $3$ for $\nu>1000$.
In the  high-precision settings stored in the file {\tt cl\_ref.pre}, $N=10$ for any value of $\nu$. These numerical values can of course be changed by the user\footnote{In \CLASS{} v2.0, the sampling parameter is fixed by the precision parameters {\tt hyper\_sampling\_curved\_high\_nu}, {\tt hyper\_sampling\_curved\_low\_nu}, {\tt hyper\_nu\_sampling\_step}, which by default are equal to 3, 6, 1000.}. 

\subsection{Radial functions\label{sec:radial}}

The radial functions $\phi_\ell^{j(m)}$, $\epsilon_\ell^{(m)}$, $\beta_\ell^{(m)}$ are given in Ref.~\cite{Hu:1997mn} in terms of the hyperspherical functions $\Phi_\ell ^\nu(\chi)$. In the code, we wrote these functions in such a way that the flat-space limit is completely transparent. We define five functions with simple flat-space limits:
\begin{align}
\genbes(\chi) &\equiv \Phi_\ell ^\nu(\chi), &\genbes(\chi) &\xrightarrow[K\rightarrow0]{} j_\ell (k (\tau_0-\tau)), \\
\dgenbes(\chi) &\equiv \frac{\sqrt{|K|}}{k} \Phi_\ell ^{\nu\prime} (\chi), &\dgenbes(\chi) &\xrightarrow[K\rightarrow0]{} j_\ell '(k (\tau_0-\tau)), \\
\ddgenbes(\chi) &\equiv \frac{|K|}{k^2} \Phi_\ell ^{\nu\prime\prime} (\chi), &\ddgenbes(\chi) &\xrightarrow[K\rightarrow0]{} j_\ell ''(k (\tau_0-\tau)), \\
\gencsck(\chi) &\equiv \frac{\sqrt{|K|}}{k} \cscK(\chi), &\gencsck(\chi) &\xrightarrow[K\rightarrow0]{} \frac{1}{k (\tau_0-\tau)}, \\
\gencotk(\chi) &\equiv \frac{\sqrt{|K|}}{k} \cotK(\chi), &\gencotk(\chi) &\xrightarrow[K\rightarrow0]{} \frac{1}{k (\tau_0-\tau)}~.
\end{align}
Note that the last two functions are related to each other through
\begin{equation}
\cotK^2(\chi) + \kk = \cscK^2(\chi) ~~~~\Rightarrow~~~~ \gencotk^2(\chi) + \frac{K}{k^2} = \gencsck^2(\chi).
\end{equation}
In Ref.~\cite{Tram:2013ima}, we defined a set of numbers $s_\ell $ which enter in the coefficients of the Boltzmann hierarchy:
\begin{equation}
s_\ell  = \sqrt{1-\frac{\ell ^2-1}{k^2}K},
\end{equation}
where $\ell $ is a positive integer. However, for the purpose of writing the radial functions in a simple form, we will extend this definition to $\ell \in \mathbb{C}$. Specifically, we need
\begin{equation}
k^2+pK =k^2\(1-\frac{-p}{k^2}K\) = k^2 s_{\sqrt{1-p}}^2,
\end{equation}
where $p$ is a positive or negative integer. We find explicitly
\begin{align}
\sqrt{k^2-K} &=k s_{\sqrt{2}}, &
\sqrt{k^2-2K} &= k s_{\sqrt{3}}, &
\sqrt{k^2-3K} &= k s_2, & 
\sqrt{k^2-4K} &= k s_{\sqrt{5}}, \nonumber \\
\sqrt{k^2+K} &=k s_0, &
\sqrt{k^2+2K} &=k s_{\ii}, &
\sqrt{k^2+3K} &=k s_{\sqrt{2}\ii}, &
\sqrt{k^2+4K} &=k s_{\sqrt{3}\ii}. \label{eq:explicitsl}
\end{align}
We can rewrite the radial functions of~\cite{Hu:1997hp} and~\cite{Hu:1997mn} in a compact and unified way,
\paragraph{Scalar modes:}
\begin{subequations}
\begin{flalign}
\qquad \phi_\ell ^{(00)} &= \genbes(\chi), \qquad\quad
\phi_\ell ^{(10)} = \dgenbes(\chi), &
\phi_\ell ^{(20)} &= \frac{1}{2 s_2} \[3 \ddgenbes(\chi) + \genbes(\chi) \], & \\
\epsilon_\ell ^{(0)} &=  \sqrt{\frac{3(\ell +2)!}{8(\ell -2)!}} \frac{1}{s_2}\gencsck^2(\chi) \genbes(\chi), &
\beta_\ell ^{(0)} &=0. &
\end{flalign}
\end{subequations}
\paragraph{Vector modes:}
\begin{subequations}
\begin{flalign}
\qquad \phi_\ell ^{(11)} &= \frac{1}{\sqrt{2}} \sqrt{\ell (\ell +1)} \frac{1}{s_0} \gencsck(\chi) \genbes (\chi), &\\
\phi_\ell ^{(21)} &= \sqrt{\frac{3}{2}} \sqrt{\ell (\ell +1)} \frac{1}{s_{\sqrt{3}} s_0}  \gencsck(\chi) \[\dgenbes(\chi) -\gencotk(\chi) \genbes(\chi) \], &\\
\epsilon_\ell ^{(1)} &= \frac{1}{2} \sqrt{(\ell -1)(\ell +2)} \frac{1}{s_{\sqrt{3}} s_0} \gencsck(\chi) \( \gencotk(\chi) \genbes(\chi) + \dgenbes(\chi) \) ,  &\\
\beta_\ell ^{(1)}	&= \frac{1}{2} \sqrt{(\ell -1)(\ell +2)} \frac{s_{\ii} }{s_{\sqrt{3}} s_0}  \gencsck(\chi) \genbes (\chi).
\end{flalign}
\end{subequations}
\paragraph{Tensor modes:}
\begin{subequations}
\begin{flalign}
\qquad \phi_\ell ^{(22)} &= \sqrt{\frac{3(\ell +2)!}{8(\ell -2)!}} \frac{1}{s_{\sqrt{2}} s_{\ii}}  \gencsck^2(\chi) \genbes (\chi), &\\
\epsilon_\ell ^{(2)}	&= \frac{1}{4} \frac{1}{ s_{\sqrt{2}} s_{\ii}} \[ \ddgenbes(\chi) + 4 \gencotk(\chi) \dgenbes(\chi) -\(s_{\sqrt{3}\ii}^2 -2\gencotk^2(\chi) \) \genbes(\chi) \], &\\
\beta_\ell ^{(2)}	&= \frac{1}{2} \frac{s_{\sqrt{2}\ii}}{s_{\sqrt{2}} s_{\ii}}  \[\dgenbes (\chi) +2\gencotk(\chi) \genbes(\chi) \]. 
\end{flalign}
\end{subequations}
In \CLASS{}, radial functions are found in exactly that form in the function {\tt transfer\_radial\_function()} inside the ``transfer'' module.

\subsection{Harmonic power spectra\label{sec:spectra}}

In flat or open space, the integral leading to the harmonic power spectra ${C_\ell^{XY}}^{(m)}$ (with $X$ referring to $T$, $E$, $B$, or the density or lensing potential of sources in a given bins, and $m=0,1,2$ for scalars, vectors, tensors) reads~\cite{Hu:1997hp,Hu:1997mn} 
\begin{equation}
{C_\ell^{XY}}^{(m)} = 4 \pi \int \frac{dk}{k}\,\,   {\Delta_\ell^X}^{(m)}(q,\tau_0) \,\, {\Delta_\ell^Y}^{(m)}(q,\tau_0) \,\, {\cal P}^{(m)}(k)~, \label{eq:spec1}
\end{equation}
where $q$ is seen as a function of $k$, and ${\cal P}^{(m)}(k)$ is the primordial spectrum of the mode $m$. In closed space, the integral is replaced by a sum over discrete wavenumbers $k_n=\sqrt{(n^2 - 1-m)K}$, corresponding to $\nu_n=q_n/\sqrt{K}=n$, with $n=3,4,5,...,\infty$.

For vector modes, there is no significant generation of CMB anisotropies through a passive mechanism, i.e. from initial conditions in the very early universe, so the vector primordial spectrum is not a relevant quantity. For scalar and tensor modes, the definition and normalisation of the primordial spectrum in a non-flat universe is presented in the literature under different forms, sometimes confusing, and usually with scarce explanations. In this section we wish to clarify this issue, to introduce the conventions used in both \CAMB{} and \CLASS{}, and to compare them with expressions appearing in the literature.

In general, each primordial spectrum refers to a given perturbation $A$ (metric, curvature, density, etc.) which can be chosen almost arbitrarily. For consistency, the transfer functions $\Delta_\ell^X(q)$ must be computed starting from a set of initial conditions for the system of cosmological perturbations such that the quantity $A$ has initially a unit value. At initial time the power spectrum $P_A(k)$ of $A$ is defined as
\begin{equation}
\langle A(\vec{k}) A^*(\vec{k}') \rangle = P_A(k) \delta(\vec{k}-\vec{k}') ~,
\end{equation}
and the dimensionless power spectrum ${\cal P}_A(k)$ follows from ${\cal P}_A(k)=\frac{k^3}{2 \pi^2} P_A(k)$. This definition is designed in such a way that the dimensionless power spectrum represents the contribution of a given logarithmic interval to any integral in $k$ space:
\begin{equation}
\forall f, \qquad \int \frac{d^3 \vec{k}}{(2\pi)^3} P_A(k) f(k) = \int \frac{dk}{k} {\cal P}_A(k) f(k)~.
\end{equation}
\paragraph{Scalar modes.}
For adiabatic initial conditions, in \CAMB{} and \CLASS{}, ${\cal P}^{(0)}(k)$ is assumed to be the curvature power spectrum ${\cal P}_{\cal R}(k)$, where ${\cal R}$ stands for the comoving curvature perturbation in the comoving gauge. Then the adiabatic transfer function must be computed starting from the initial condition ${\cal R} \longrightarrow 1$ when $k \tau \longrightarrow 0$. This corresponds to $\eta \longrightarrow 1$ in the synchronous gauge, still using the notations of Ref.~\cite{Ma:1995ey} for synchronous metric perturbations. The equivalent initial condition in the Newtonian gauge is obtained from a simple gauge transformation. Hence the scalar harmonic spectrum is given by
\begin{equation}
{C_\ell^{XY}}^{(0)} = 4 \pi \int \frac{dk}{k} \,\, {\Delta_\ell^{X}}^{(0)}(q,\tau_0) \,\, {\Delta_\ell^{Y}}^{(0)}(q,\tau_0) \,\, {\cal P}_{\cal R}(k)~,
\label{eq:spec2}
\end{equation}
or by the corresponding sum in a closed universe. In the case of mixed initial condition (adiabatic plus isocurvature), equation~(\ref{eq:spec2}) needs to be replaced by a sum over each pair of initial conditions, involving auto- and cross-correlation primordial spectra ${\cal P}_{\cal R \cal R}(k)$, ${\cal P}_{\cal R \cal I}(k)$, ${\cal P}_{\cal I \cal I}(k)$. 

Lyth \& Stewart \cite{Lyth:1990dh} showed that  the dimensionless scalar spectrum of primordial metric fluctuations ${\cal P}_{\phi}(k)$ generated by single-field inflation in an open universe is scale-invariant in the limit of negligible slow-roll parameters, i.e. for $n_s=1$. This result also holds in the closed case \cite{Starobinsky:1996ek}. Since ${\cal R}= \frac{3}{2} \phi$ on super-Hubble scale and during radiation domination, this corresponds to a constant curvature spectrum, ${\cal P}_{\cal R}\equiv A_s$. To account for deviations from this limit, it is customary to introduce a power-law primordial spectrum
\begin{equation}
{\cal P}_{\cal R}(k) = A_s k^{n_s-1},
\label{eq:scalprispec}
\end{equation}
where the scalar tilt $n_s$ can be related to the first-order slow-roll parameters. 
In the literature, the integral of equation~(\ref{eq:spec2}) (or the corresponding sum in a closed universe) is often written in terms of $q$ or $\nu$:
\begin{eqnarray}
{C_l^{XY}}^{(0)} 
%&=& 4 \pi \int \frac{dk}{k} \Delta_{lX}^{(0)}(q,\tau_0) \Delta_{lY}^{(0)}(q,\tau_0) {\cal P}_{\cal R}(k) \\
%&=& 4 \pi \int \frac{q\,dq}{k^2} \Delta_{lX}^{(0)}(q,\tau_0) \Delta_{lY}^{(0)}(q,\tau_0) {\cal P}_{\cal R}(k) \nonumber \\
&=& 4 \pi \int dq \, q^ 2 {\Delta_\ell^X}^{(0)}(q,\tau_0) {\Delta_\ell^Y}^{(0)}(q,\tau_0) \frac{{\cal P}_{\cal R}(k)}{q\left(q^2- K\right)} \label{eq:spec3}\\
&=& 4 \pi \int d\nu \, \nu^ 2 {\Delta_\ell^X}^{(0)}(q,\tau_0) {\Delta_\ell^Y}^{(0)}(q,\tau_0) \frac{{\cal P}_{\cal R}(k)}{\nu\left(\nu^2-  \hat{K} \right)}~,\label{eq:spec4}
\end{eqnarray}
where we used $q^2 = k^2+K$ and $k\,dk = q\,dq$. The literature also often refers to a dimensionless primordial spectrum in $q$-space that we will denote $\widetilde{\cal P}_{\cal R}(q)$. Since the dimensionless spectrum refers to the contribution of a given logarithmic interval, it is natural to define $\widetilde{\cal P}_{\cal R}(q)$ as:
\begin{equation}
\frac{dk}{k}  {\cal P}_{\cal R}(k) \equiv \frac{dq}{q}  \widetilde{\cal P}_{\cal R}(q)
\qquad \Longrightarrow \qquad
\widetilde{\cal P}_{\cal R}(q) = \frac{q^2}{\left(q^2- K\right)}{\cal P}_{\cal R}(k)~. \label{eq:corpkpq}
\end{equation}
\paragraph{Tensor modes.}
The perturbed FLRW metric $g_{\mu \nu} = a^2 \left( \gamma_{\mu \nu} + h_{\mu \nu}\right)$
(with $\gamma_{00}=-1$) features gravitational waves in the traceless transverse part of $h_{ij}$. For each wavelength and direction of propagation of the waves, $h_{ij}$ contains two independent degrees of freedom (called the two polarisation states). In a Friedmann-Lema\^{\i}tre universe, the evolution equation and initial power spectrum of these two states are identical, and independent of the direction of propagation. Hence the statistical properties of gravitational waves can be inferred from a unique power spectrum
${\cal P}_H(k)$ and transfer function $H(\tau, k)$. In standard notations, $H$ is normalized in such a way that it obeys the evolution equation (see e.g.~\cite{Hu:1997mn})
\begin{equation}
H'' + 2\frac{a^\prime}{a} H + \left(k^2+2K\right) H =  8 \pi G a^2  \sum_{i=\gamma,\nu} 
p_i \, \pi^{(2)}_i ~,
\end{equation}
where the anisotropic stress $\pi^{(2)}_i$ is related to the coefficients of the optimal Bolztmann hierarchy of Ref.~\cite{Tram:2013ima} through
\begin{equation}
\pi^{(2)}_i = -4 \sqrt{6} \left(\frac{1}{15} F^{(2)}_{i 0} + \frac{2}{21} F^{(2)}_{i 2} + \frac{1}{35} F^{(2)}_{i 4}\right)~,
\end{equation}
and the index $i$ runs over relativistic photons and neutrinos\footnote{The contribution from massive neutrinos has not yet been implemented.} (non-relativistic particles have a negligible anisotropic stress, and do not contribute to the source term). The tensor harmonic spectra follows from
\begin{eqnarray}
{C_\ell^{XY}}^{(2)} &=& 4 \pi \int \frac{dk}{k} \,\, {\Delta_\ell^X}^{(2)}(q,\tau_0) \,\, {\Delta_\ell^Y}^{(2)}(q,\tau_0) \,\, {\cal P}_{H}(k) \\
&=& 4 \pi \int \frac{dq}{q} \,\, {\Delta_\ell^X}^{(2)}(q,\tau_0) \,\, {\Delta_l^Y}^{(2)}(q,\tau_0) \,\, \widetilde{\cal P}_{H}(q)~,
\end{eqnarray}
or the equivalent sum of discrete $k$ modes in  the case of a closed universe. The relation between ${\cal P}_{H}(k)$ and
$\widetilde{\cal P}_{H}(q)$ is now given by
\begin{equation}
\frac{dk}{k}  {\cal P}_H(k) \equiv \frac{dq}{q}  \widetilde{\cal P}_H(q)
\qquad \Longrightarrow \qquad
\widetilde{\cal P}_H(q) = \frac{q^2}{\left(q^2- 3 K\right)}{\cal P}_H(k)~. \label{eq:corpkhpq}
\end{equation}
It was shown by \cite{Bucher:1997xs} (see also \cite{Hu:1997ws}) that during slow-roll inflation in a bubble with negative curvature, and in the limit of negligible slow-roll parameters, the minimal primordial spectrum of gravitational waves is proportional to
\begin{equation}
 {\cal P}_{H}(k) \propto \frac{k^2(k^2-K)}{(k^2+3K)(k^2+2K)} \, \mathrm{tanh} 
 \left( \frac{\pi}{2} \sqrt{\frac{k^2+3K}{K}}\right)~, \qquad (K < 0).
 \end{equation}
 The hyperbolic tangent cut-off  accounts for the suppression of wavelengths which are large compared to the radius of curvature of the universe. In a closed universe the same result applies but without the cut-off. Hence it is convenient to define
 \begin{equation}
 t_K(k) =  \left\{
 \begin{tabular}{ll}
 $\mathrm{tanh} \left( \frac{\pi}{2} \sqrt{\frac{k^2+3K}{K}} \right) = \mathrm{tanh} \left( \frac{\pi \nu}{2} \right)$ & $\mathrm{if}~K < 0$~,\\
$1$&$\mathrm{if}~K\geq 0$~.
 \end{tabular}
 \right.
 \end{equation}
The normalisation of the spectrum is set by a parameter $A_t$, and deviations from the slow-roll limit are encoded at first order in a power-law $k^{n_t}$, where the tensor tilt $n_t$ relates to the first slow-roll parameter. The usual normalisation convention for $A_t$ is such that for single-field inflation and in the slow-roll limit, the tensor-to-scalar ratio $r \equiv A_t/A_s$ relates to the tensor index through
\begin{equation}
r \equiv \frac{A_t}{A_s} = 16 \epsilon = -8 n_t~.
\end{equation}
A careful computation shows that this definition is compatible with
\begin{equation}
{\cal P}_H(k) = \frac{A_t}{6} k^{n_t} \frac{k^2(k^2-K)}{(k^2+3K)(k^2+2K)} t_K(k)~, 
\end{equation}
which corresponds, using (\ref{eq:corpkhpq}), to
\begin{equation}
\widetilde{\cal P}_H(q) = \frac{A_t}{6} k^{n_t} \frac{(k^2-K)}{(k^2+2K)} t_K(k) = \frac{A_t}{6} k^{n_t} \frac{(q^2-4K)}{(q^2-K)} t_K(q) = \frac{A_t}{6} k^{n_t} \frac{(\nu^2-4\hat{K})}{(\nu^2-\hat{K})} t_K(\nu)~. 
\end{equation}
\paragraph{Comparison with other references.}
Ref.~\cite{Hu:1997mn} refers to the primodial spectra of metric fluctuations $\phi$ and tensor fluctuations $H$. Following our notations, we infer from previous results that in the slow-roll limit $n_s=1$ and $n_t=0$, these spectra should be given by
\begin{eqnarray}
P_\phi(q) &=& \frac{2 \pi^2}{q^3} \widetilde{\cal P}_\phi (q) \propto \frac{1}{q(q^2-K)} \propto \frac{1}{\nu\left(\nu^2-  \hat{K} \right)}~,\\
P_H(q) &=& \frac{2 \pi^2}{q^3} \widetilde{\cal P}_H (q) \propto \frac{q^2-4K}{q^3(q^2-K)} \propto  \frac{(\nu^2-4\hat{K})}{\nu^3(\nu^2-\hat{K})}~t_K(\nu)~.
\end{eqnarray}
This coincides exactly with equation~(44) of \cite{Hu:1997mn} for $\hat{K}=-1$. In Lyth \& Stewart \cite{Lyth:1990dh}, the final result for scalar modes is expressed in terms of the dimensionless density power spectrum $P_\delta$. Using the Poisson equation
\begin{equation}
4 \pi G a^2 \bar{\rho}_\mathrm{tot} \delta_\mathrm{tot} = - s_2^2 k^2 \phi = - (k^2-3K) \phi = - (q^2-4K) \phi~,
\end{equation}
we see that in the slow-roll limit where $n_s=1$,
\begin{equation}
P_\delta(q) \propto (q^2-4K)^2 P_\phi(q) \propto  \frac{(q^2-4K)^2}{q(q^2-K)} \propto \frac{(\nu^2-4\hat{K})^2}{\nu(\nu^2-\hat{K})} ~. \label{eq:densprispec2}
\end{equation}
This result coincides with \cite{Lyth:1990dh} for $\hat{K}=-1$.
Seljak et al. \cite{Zaldarriaga:1997va} use equation~(\ref{eq:densprispec2}) to define their scalar primordial spectrum. By doing so, they implicitly assume that the transfer functions ${\Delta_\ell^X}^{(0)}(k,\tau_\mathrm{ini})$ are normalised at initial time to $\delta_\mathrm{tot}=1$ instead of the more conventional choices $\phi=1$, or ${\cal R}=1$ like in \CAMB{} and \CLASS{}.

\section{Flat rescaling approximation\label{sec:approx}}

In a curved universe, the comoving angular diameter distance to the last scattering surface reads
\begin{equation}
r_A^\mathrm{rec} = \frac{d_A^\mathrm{rec}}{a^\mathrm{rec}} = \frac{1}{\sqrt{|K|}} \sin_K \left(\sqrt{|K|} (\tau_0-\tau_\mathrm{rec})\right)~,
\end{equation}
where $\tau_\mathrm{rec}$ is the conformal time at recombination. This means that features on the last scattering surface are shifted in angular space by
\begin{equation}
\alpha = \frac{r_A^\mathrm{rec}}{(\tau_0-\tau_\mathrm{rec})} \label{eq:alpha}
\end{equation}
with respect to a universe with the same conformal age and recombination time\footnote{This ratio $\alpha$ is computed by \CLASS{} in the {\tt thermodynamics} module and called {\tt angular\_rescaling}. It is used in many places in the code since all characteristic quantities in multipole space (sampling step sizes, values at which a given approximation should be switched on) are systematically scaled by $\alpha$.}, as clearly seen in the left panel of figure~\ref{fig:decomposition}. This suggests various possible approximation schemes, like computing the CMB spectra $C_l^{XY}$ in flat space and rescaling them horizontally by $\alpha$, or replacing the hyperspherical Bessel functions $\Phi_l^\nu(\chi)$ by $j_l(\alpha \nu \chi)$. Such schemes were already attempted by~\cite{Zaldarriaga:1997va} and found to be very inaccurate. Still, some good approximation must exist, to account for the fact that when the curvature is small, the CMB physics is almost the same as in flat space, except for a small impact of the curvature $K$ on the evolution of the smallest wavenumbers $k \sim K$, and for a small change in the angular diameter distance to recombination. The modes with $k \gg K$, i.e. $\nu \gg1$, only experience the latter effect. 
\subsection{Approximating hyperspherical Bessel functions \label{sec:ahbf}}
We investigate an approximation scheme which is supposed to be valid above a critical value $\nu_*$. For large $\nu$'s, one notices that the hyperspherical bessel functions are very close to their flat spherical counterpart with a linearly rescaled argument. Our goal is then to use an approximation of the type
\begin{equation}
\Phi_l^\nu(\chi) \simeq A_\nu^l(\chi) j_l(\gamma \nu \chi) \qquad \mathrm{if}~\nu\geq\nu_* ~,\label{eq:flat-approx} 
\end{equation}
where the oscillatory part is described by the flat Bessel functions, while $A$ is a smooth function of $\chi$. 

Ideally, we would like our approximation scheme to reach its maximum precision for $\chi$ values close to the first peak of the Bessel function, since this is the region contributing most to the integrals of equations~(\ref{eq:transT}, \ref{eq:transB}). The position and amplitude of the first peak is difficult to estimate analytically. It is more convenient to match the exact and approximate functions at the turning point, i.e. at the first point where the second derivative vanishes, located just before the first peak. Its position is given exactly by the equation
\begin{equation}
\sinK \chi_\mathrm{tp} =  \frac{\sqrt{l(l+1)}}{\nu} ~.
\end{equation}
Hence the rescaling factor $\gamma$ can be defined as the ratio of the turning points for the hyperspherical and spherical Bessel functions:
\begin{equation}
\gamma = \frac{\sqrt{l(l+1)}}{\nu \,\, \mathrm{arcsin}_K\left( \frac{\sqrt{l(l+1)}}{\nu}\right)}~. \label{rescaling_argument}
\end{equation}
The amplitude of the hyperspherical Bessel function compared to the spherical one around the turning point  can
 be estimated from the ratio of their respective WKB approximations~\cite{Kosowsky:1998nc,Tram:2013xpa} in the limit $\chi \rightarrow \chi
_\text{tp}$. By defining $Q(\chi)\equiv \cscK^2 \chi - \frac{\nu^2}{l(l+1)}$ and expanding $Q(\chi) \simeq Q'(\chi_\text{tp}) (\chi - \chi_\text{tp} 
)$ around $\chi_\text{tp}$, equation~\eqref{eq:u_equation} becomes the Airy equation. Evaluating the solutions at $\chi = \chi_\text{tp}$ we find
\begin{equation}
\Phi_l^\nu (\chi_\text{tp}) \sim Q'(\chi_\text{tp}) ^{-\frac{1}{6}} = \( \frac{-2 \sinK \chi_\text{tp} \cotK \chi_\text{tp}}{\sinK^3 \chi_\text{tp}} \)^{-\frac{1}{6}} = \( \frac{-2 \sinK \chi_\text{tp} \cotK \chi_\text{tp}}{\[\frac{l(l+1)}{\nu^2}\]^{\frac{3}{2}}} \)^{-\frac{1}{
6}},
\end{equation}
so the difference in amplitude at the turning point is due to the product: $\sinK \chi_\text{tp} \cotK \chi_\text{tp}$. Since $\sinK \chi \cotK
 \chi = 1$ for $\kk=0$, the ratio of amplitudes at the turning point is exactly
\begin{align}
A_\nu^l(\chi_\text{tp}) \equiv \frac{\Phi_l^{\nu,\text{ \text{curved}}} (\chi_\text{tp})}{\Phi_l^{\nu,\text{ \text{flat}}} (\chi_\text{tp})} 
&= \( \sinK \chi_\text{tp} \cotK \chi_\text{tp} \) ^{-\frac{1}{6}} = \left\{
\begin{array}{ll}
\(\cosh \asinh \[ \frac{\sqrt{l(l+1)}}{\nu} \] \) ^{-\frac{1}{6}}, & \kk =-1, \\
\(\cos \asin \( \frac{\sqrt{l(l+1)}}{\nu} \] \) ^{-\frac{1}{6}},  & \kk =1,
\end{array}
\right. \nonumber \\
&=  \[ 1 -\kk \frac{l(l+1)}{\nu^2} \]^{-\frac{1}{12}}.
\end{align}
Using constant rescaling factors $\gamma$ and $A_\nu^l(\chi_\text{tp})$ in equation~(\ref{eq:flat-approx}) already provides a good approximation, but not good enough for our purpose, since such a scheme leads to percent level errors in the $C_\ell$'s. The reason is that the ratio $A_\nu^l(\chi_\text{tp}) $ defined above gives a very good approximation of the amplitude of the first peak, but overestimates (underestimates) the amplitude of the subsequent peaks for $K<0$ ($K>0$). We know from the WKB approximation that in the limit $\chi \longrightarrow \infty$, the hyperspherical to spherical peak ratio approaches $\chi / \sin_K(\chi)$, but the CMB is mainly sensitive to the first hundred peaks or so, which are still far from this asymptotic regime. We find numerically that a quadratic rescaling formula of the type\footnote{We deduced the functional dependence on $l$ and $\nu$ by fitting second order polynomials to the ratios of peaks on a $(l,\nu)$-grid. We then observed that the coefficients seemed to depend mostly on the angle $\beta = \atan(l/\nu)$ in this plane.}
\begin{equation}
A_\nu^l(\chi) = A_\nu^l(\chi_\text{tp}) \left[ 1 + a \,\, \atan\left({l}/{\nu}\right) \,\, (\chi-\chi_\text{tp}) + b \,\, \left(\atan\left({l}/{\nu}\right)\right)^2 \,\, (\chi-\chi_\text{tp})^2 \right]~, \label{rescale_function}
\end{equation}
with
\begin{eqnarray}
a=0.34, & b= 2.0 & (K>0)~,\nonumber \\
a=-0.38, & b=0.4 & (K<0)~,
\end{eqnarray}
leads to 0.1\% precision on the $C_\ell$'s  with $\nu_*=4000$, as discussed later and shown in Figure~\ref{fig:flatapprox}. In the flat limit, $K\rightarrow 0$, we have $\nu \rightarrow \infty$ and consequently $\atan(l/\nu)\rightarrow 0$. So $A_\nu^l(\chi)\rightarrow 1$ in the flat limit as it should.

At this point we would like to emphasise that we are not trying to interpolate transfer functions or power spectra as a function of the cosmological parameter $K$. Such a scheme would be most effective when implemented at the level of a parameter extraction code such as Pico~\cite{Fendt:2006uh}. What we are doing is just approximating one special function with another which is easy to compute in a region of parameters where the two must be similar. This is in fact similar to the uniform WKB approximation where the hyperspherical Bessel functions are related to the easily computable Airy function.
\subsection{Implementation of the flat approximation}
\begin{figure}\label{fig:flatapprox}%
\includegraphics[width=0.5\columnwidth]{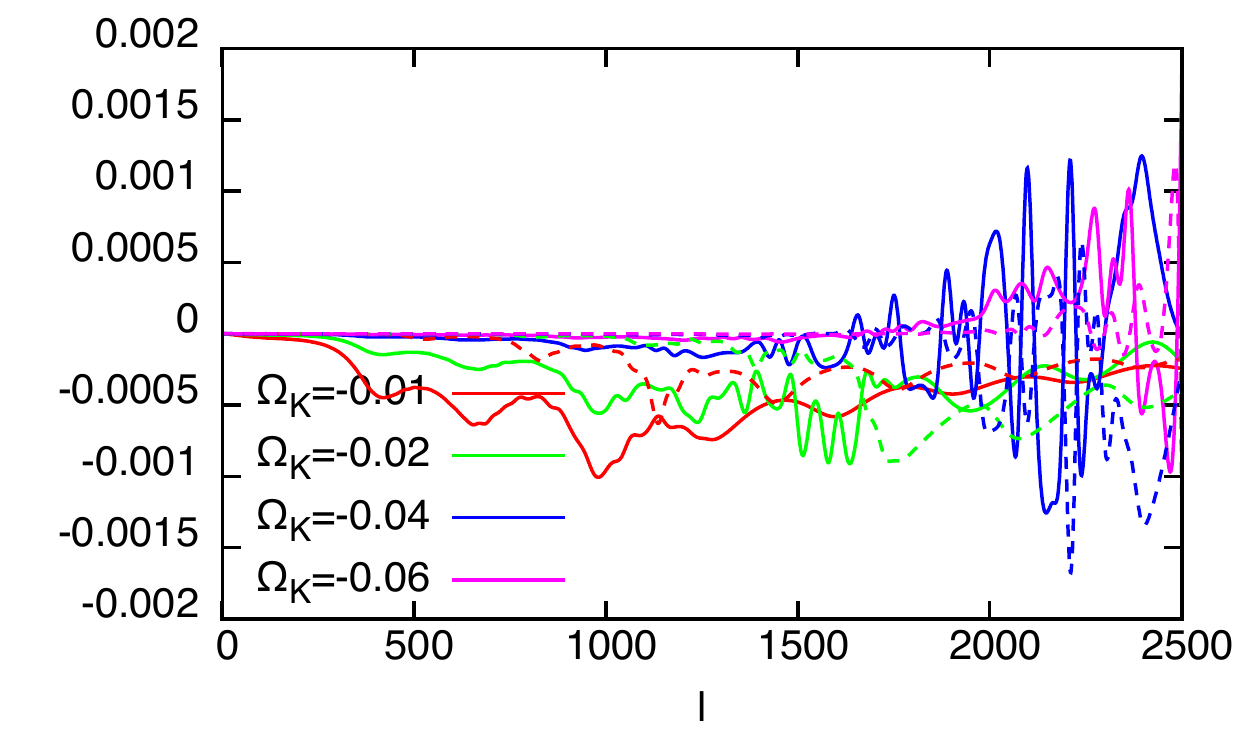}%
\includegraphics[width=0.5\columnwidth]{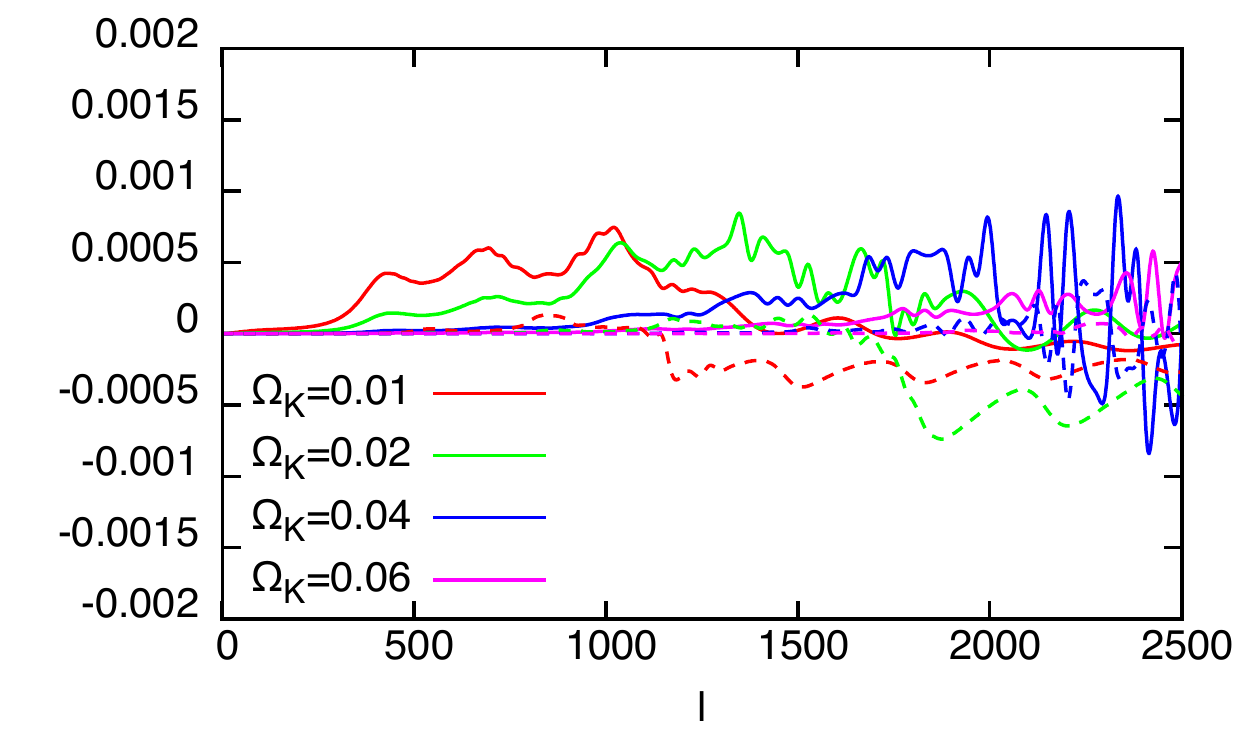}%
\caption{Relative difference between scalar unlensed $C_\ell$'s computed with and without the flat rescaling approximation for $\nu_*=4000$. Temperature $C_\ell$'s are displayed in solid lines and E-type polarisation is shown in dashed lines. \emph{Left panel:} Positive curvature. \emph{Right panel:} Negative curvature. Different colours correspond to different values of $\Omega_K$ picked up in the range where the approximation is the most harmful: for smaller $|\Omega_K|$'s the approximation is very accurate at every $l$, while for larger $|\Omega_K|$'s it introduces a significant error  only for multipoles $l > 2500$. All calculations have been performed using the reference settings of the file {\tt cl\_ref.pre} for all precision parameters but $\nu_*$.}%
\end{figure}

In \CLASS{} v2.0, the {\tt transfer} module first computes the flat spherical Bessel functions and stores them in memory. It then loops over growing values of $\nu$. For $\nu<\nu_*$, Bessel functions $\Phi_l^\nu(\chi)$ are recomputed at the start of each new loop. For $\nu>\nu_*$ this computation is switched off. The code interpolates from the array of flat Bessel function with a rescaled argument, using the rescaling factor of equation~(\ref{rescaling_argument}), and multiplies the result by the rescaling function of equation~(\ref{rescale_function}). For simplicity, $n$-th derivatives of hyperspherical Bessel functions are approximated as
\begin{equation}
{\Phi_l^{\nu}}^{(n)}(\chi) \simeq A_\nu^l(\chi)\, (\gamma \nu)^n \, j_l^{(n)}(\gamma \nu \chi) \qquad \mathrm{if}~\nu\geq\nu_* ~,\label{eq:flat-approx-derivs} 
\end{equation}
neglecting derivatives of the smoothly varying $A_\nu^l$ function. Models with a smaller curvature have a larger fraction of their $\nu$ values above $\nu_*$, so the execution time of the code tends towards that of flat models in the limit $K\longrightarrow0$. For very small $K$, all Bessel functions are replaced by their flat rescaled approximation, with rescaling factors $\gamma$ and $A_\nu^l(\chi)$ tending towards one. This ensures that $C_\ll^{XY}$ is perfectly continuous with respect to $K$ across the special point $K=0$. In the default version of \CLASS{} v2.0, the approximation is switched on at $\nu_*=4000$\footnote{In the code, $\nu_*$ is called {\tt hyper\_flat\_approximation\_nu}.}. In the  high-precision settings stored in the file {\tt cl\_ref.pre}, this number is pushed up to $10^6$. In Figure~\ref{fig:flatapprox}, we show the ratio of $C_\ell^{TT}$'s and  $C_\ell^{EE}$'s computed with and without this approximation. 

The error introduced by the approximation goes to 0 when $|K| \longrightarrow \infty$, because it is never used (all $\nu$ values being below $\nu_*$), and we checked that the same is true in the limit $|K| \longrightarrow 0$, in which the approximate $\Phi_l^\nu$ tends towards the real ones. For intermediate values of $|K|$, the approximation affects $C_\ell^{XY}$'s above some value $l_*(\nu_*)$ growing with $\nu_*$. Above this multipole, the amplitude of the error also grows with  $\nu_*$.
For the choice $\nu_*=4000$ and positive curvature, models with $|\Omega_K| < 0.1$ are affected at almost any $l$, but by a negligible amount. For $|\Omega_K|>0.06$, the approximation is harmless because values with $2<l<2500$ are almost unaffected by the approximation (used only in the tail of the integrals over time). Larger $l$'s are more affected, but they are not constrained very accurately by CMB data. In the intermediate range $0.01<|\Omega_K|<0.06$, the error in the range $2<l<2500$ is maximal. Figure~\ref{fig:flatapprox} shows that this error peaks around 0.1\% for temperature and E-type polarisation $C_\ell$'s. This is true for scalar spectra: tensor spectra are affected roughly at the same level, and no high-precision measurements of tensor $C_\ell$'s is foreseeable at the moment. 

We conclude that the setting $\nu_*=4000$ is sufficient for fitting Planck data, but the user is free to increase $\nu_*$ slightly in order to decrease the maximal error. If necessary, it would also be possible to increase the accuracy of the rescaling function of equation~(\ref{rescale_function}).
\subsection{Additional approximations and performance}
In this section we have described the most important new approximation which was introduced for non-zero curvature. Two additional approximations have been introduced in \CLASS{} v2.0 and they are described in appendix~\ref{sec:time_cut} and~\ref{sec:multipole_cut}. In Table~\ref{tab:performance}, we show the impact of all three approximations on the performance of the code. In previous figures, for all precision parameters not related to these three approximations, we adopted reference settings; on the contrary, for this performance test, we adopted default precision settings. All input parameters were the same as in the input file {\tt explanatory.ini} distributed with \CLASS{}, except $\Omega_k$ (switched to non-zero values in the last six models) and ${\tt modes}$ (switched to ``scalars plus tensors'' in the second model). The reported numbers account for the number of CPU cycles, rescaled in such way to give approximately the real time in seconds for a single-core run on a 2.3GHz processor. On the laptop used to perform this test, the actual execution time was approximately eight times smaller than these numbers, due to a combination of multi-threading and turbo-boost. We see that the time cut and multipole cut approximations are both important, especially in the flat case. As expected, in the non-flat case, the execution time is much larger in absence of any approximation, due to the computation of hyperspherical Bessel function. This time can be considerably reduced thanks to the flat rescaling approximation, especially for small $|\Omega_k|$. 

\begin{table}\label{tab:performance}
\begin{center}
\begin{tabular}{|l|cccccccc|}
\hline
$\Omega_k$   & 0      & 0     & $10^{-3}$ & $10^{-2}$ & $10^{-1}$ & $-10^{-3}$ & $-10^{-2}$ & $-10^{-1}$ \\
modes             & s      & s+t   & s & s & s & s & s & s \\
\hline
no approx.       & 18.0 & 28.3 & 76.6       & 78.7 & 63.8 & 95.1 & 96.7 & 163.7 \\
time cut           & 14.2 & 24.3 & 71.3       & 71.8 & 59.5 & 86.0 & 89.6 & 153.2 \\
+ $l$ cut          & 5.7   & 10.9 & 64.2       & 65.5 & 54.6 & 78.8 & 78.4 & 128.3 \\
+ flat rescaling & 5.7  & 10.9 & 18.9       & 17.6 & 32.5 & 20.8 & 21.2 & 45.4 \\
\hline
\end{tabular}
\end{center}
\caption{Execution time of \CLASS{}, for different input value of $\Omega_k$. The numbers actually reflect the number of CPU cycles, rescaled in such way to give approximately the real time in seconds for a single-core run on a 2.3GHz processor. As explained in the text, all input parameters were fixed by the file {\tt explanatory.ini}, except for  $\Omega_k$. Tensor spectra were requested only in the second column. In the first line, the three approximation schemes discussed in this section were switched off. In the next three lines, they were progressively restored. Hence the last line corresponds to the default performances of the code.}
\end{table}

\section{Accuracy and comparison with other codes\label{sec:accuracy}}

\subsection{Establishing reference precision settings and power spectra\label{sec:precision flat}}

In Ref.~\cite{Lesgourgues:2011rg}, we compared reference precision settings for \CLASS{} v1.0 and \CAMB{} (version of January 2011) for a flat minimal $\Lambda$CDM model, and we found that the temperature spectra agreed at the 0.01\% level (0.02\% for polarisation). This result extends to all ``reasonable'' $\Lambda$CDM models: there is nothing in the codes that could cause a jump in the precision when the cosmological parameters are varied over the range allowed by current cosmological data. This good agreement strongly suggests that modern Boltzmann codes are accurate at such a level. Both codes could of course be wrong if the underlying model was not correct. For instance, one could imagine that the recombination model was wrong. However, the good agreement between {\sc RECFAST}, {\sc HyRec} and {\sc CosmoRec}, all embedded within \CLASS{} and/or \CAMB{}, suggests that standard recombination physics is indeed well understood, and modelled at the accuracy level required by cosmological data. What the comparison really shows is that if the underlying physical model is correct, the codes do not introduce  a numerical error (related to numerical methods, discretisation, approximations, etc.) larger than about 0.01\%\footnote{Since the two codes are fully independent (apart from the underlying equations and the recombination modules), a very unusual coincidence would be needed to hide any larger error.}. This result is not trivial, given the high complexity of these codes.

In \cite{Lesgourgues:2011rg}, the reference settings of \CLASS{} v1.0 were obtained by varying {\it all} precision parameters\footnote{This is made easier by the fact that these parameters are all grouped inside a single structure, and set in the same place. No numbers referring to a choice of precision ever appear in the bulk of the code. Hence it is not conceivable that some precision parameters have been forgotten when doing the tuning.}, up to the point at which a typical variation of these parameters by a factor two induces less than $10^{-3}$\% variations in the $C_\ell$'s in the range $2 \leq \ell \leq 2500$. Since then, \CLASS{} has evolved, even in the flat case. Versions v1.1 to 1.7 introduced very minor changes in the $C_\ell$ computation, and we checked that their respective reference spectra remained stable at the $10^{-3}$\% level. However version 2.0 introduces a new scheme for the decomposition of sources and radial functions (already described in section~\ref{sec:sources} and \ref{sec:radial}), a new algorithm for the computation of all spherical and hyperspherical Bessel functions (described in section~\ref{sec:hyper}), and finally a new scheme for the sampling in wavenumber space (documented within the code). Hence it is important to check whether the previous 0.01\% level agreement with \CAMB{} still holds. To do so, we  established a new set of reference accuracy parameters, identical to the ones of previous versions, except for the new parameters introduced in version 2.0. For these new parameters, we again established the convergence of the code at the $10^{-3}$\% level. The resulting set of parameters is contained in the file {\tt cl\_ref.pre}, delivered together with the code. 

In figure~\ref{fig:accuracy_flat}, we show the difference between the old and new reference spectra, and conclude that the new scheme impacts the reference model at the 0.01\% level for temperature, and $0.02$\% for polarisation. Since this matches the maximum accuracy claimed for the \CLASS{} and \CAMB{} codes, we conclude that \CLASS{} v2.0 achieves the same precision as earlier versions. This is important, because the source decomposition and sampling schemes are so different in \CLASS{} v1.x and v2.0 that, together with \CAMB{}, they can almost be seen as three independent Boltzmann codes. Hence the present comparison makes our claim of a 0.01\% maximum accuracy even stronger than before.

\subsection{Accuracy of \CLASS{} and \CAMB{} in flat $\Lambda$CDM models}

\begin{figure}\label{fig:accuracy_flat}%
\includegraphics[width=0.5\columnwidth]{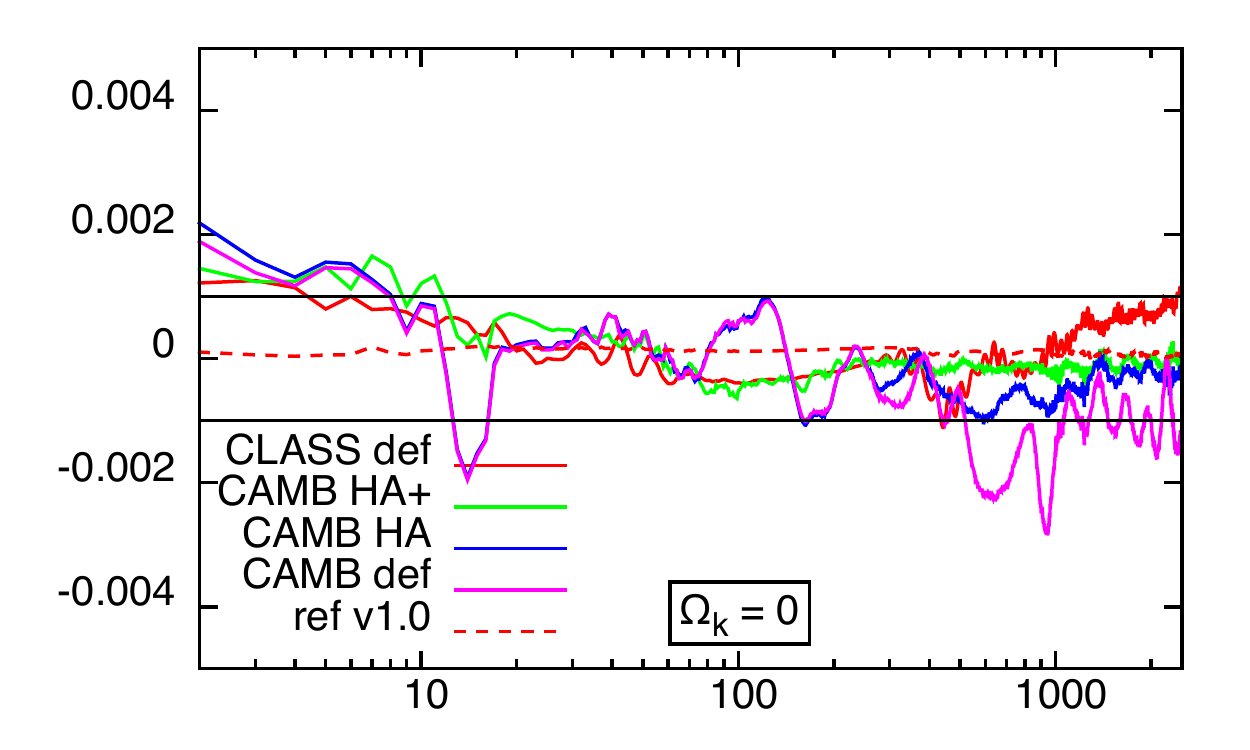}%
\includegraphics[width=0.5\columnwidth]{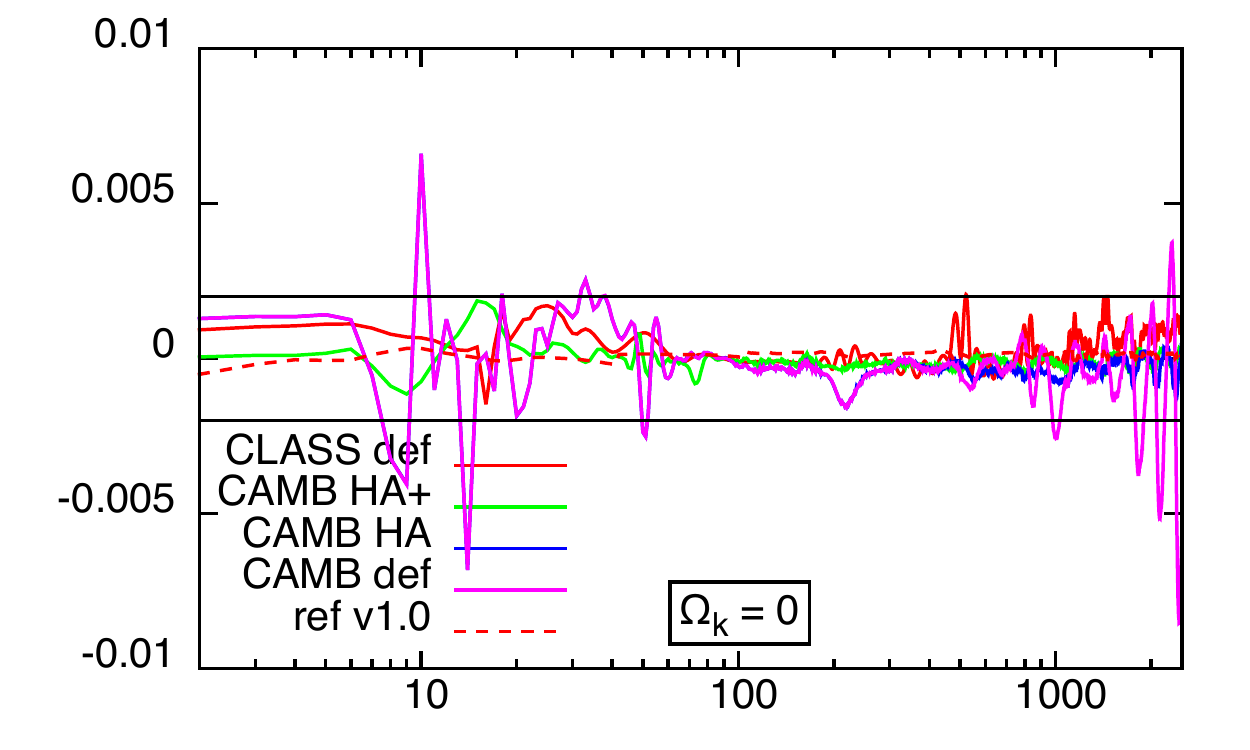}%
\caption{Accuracy of \CLASS{} and \CAMB{} in a minimal flat $\Lambda$CDM model. All spectra for unlensed temperature (left) and E-type polarisation (right) are compared to the \CLASS{} reference spectra, argued to be accurate at the 0.01\% level (see text). For \CLASS{}, we show the reference settings of v1.0, known to agree at the 0.01\% level with the high-precision limit of \CAMB{}, and the default settings of v2.0. For \CAMB{}, we show three precision settings: default, ``high accuracy'' (HA), and ``high accuracy'' plus accuracy boost parameters set to 2 (HA+).}%
\end{figure}

Having a reference model in the flat case, we degrade the precision of the new precision parameters introduced in \CLASS{} v2.0, in order to speed up the code while still achieving 0.1\% precision on scalar temperature and 0.2\% on scalar polarisation. These default settings are used when one runs the code without passing any precision parameters in input. Figure~\ref{fig:accuracy_flat} shows the resulting error for a particular minimal flat $\Lambda$CDM model. 

In Ref.~\cite{Lesgourgues:2011rg}, the reference precision settings of \CAMB{} (version of January 2011) were obtained, first, by pushing the accuracy boost parameter up to very large values, and second, by decreasing by hand the sampling step size of a few functions ($j_\ell(x)$ as a function of $x$, and ionisation fraction as a functions of $z$). Since then, the accuracy of \CAMB{} has increased: these two samplings have been improved, and the code now comes with a ``high accuracy'' flag, aimed roughly at 0.1\% precision on scalar $C_\ell$'s. Hence it is interesting to compare the spectra obtained with the latest version of \CAMB{} (from November 2013) with our reference model\footnote{All these test presented in this work rely on this version of \CAMB.}. Figure~\ref{fig:accuracy_flat} shows the accuracy achieved with default \CAMB{} precision, when using the ``high accuracy'' flag, and by using the ``high accuracy'' while also increasing the three ``accuracy boost parameters'' from 1 to 2. In the first case, the error can be as large as 0.3\% for scalar temperature, or 0.8\% for scalar polarisation. The ``high accuracy'' setting does not reduce the error for $\ell < 20$\footnote{The spike at $l=16$ had already been observed in \cite{Lesgourgues:2011rg} and has not disappeared. It might be due to a jump in the precision between $\ell=16$ and $17$, since the value {\tt llmax} is fixed to 17 in \CAMB's {\tt DoSourceIntegration} routine (A. Lewis, 2011, private communication).}, but this is harmless, due to cosmic variance. More importantly, it does keep the error at the 0.1\% level for larger $\ell$'s, making it comparable to \CLASS{} with default precision. Finally, with much higher precision settings (all ``accuracy boost parameters'' set to 2) the error decreases to about 0.05\%. We did not push the comparison further, relying on the conclusion of \cite{Lesgourgues:2011rg} that in the flat model, the agreement can be pushed to the 0.01\% level.

\begin{figure}\label{fig:accuracy_closed}%
\includegraphics[width=0.5\columnwidth]{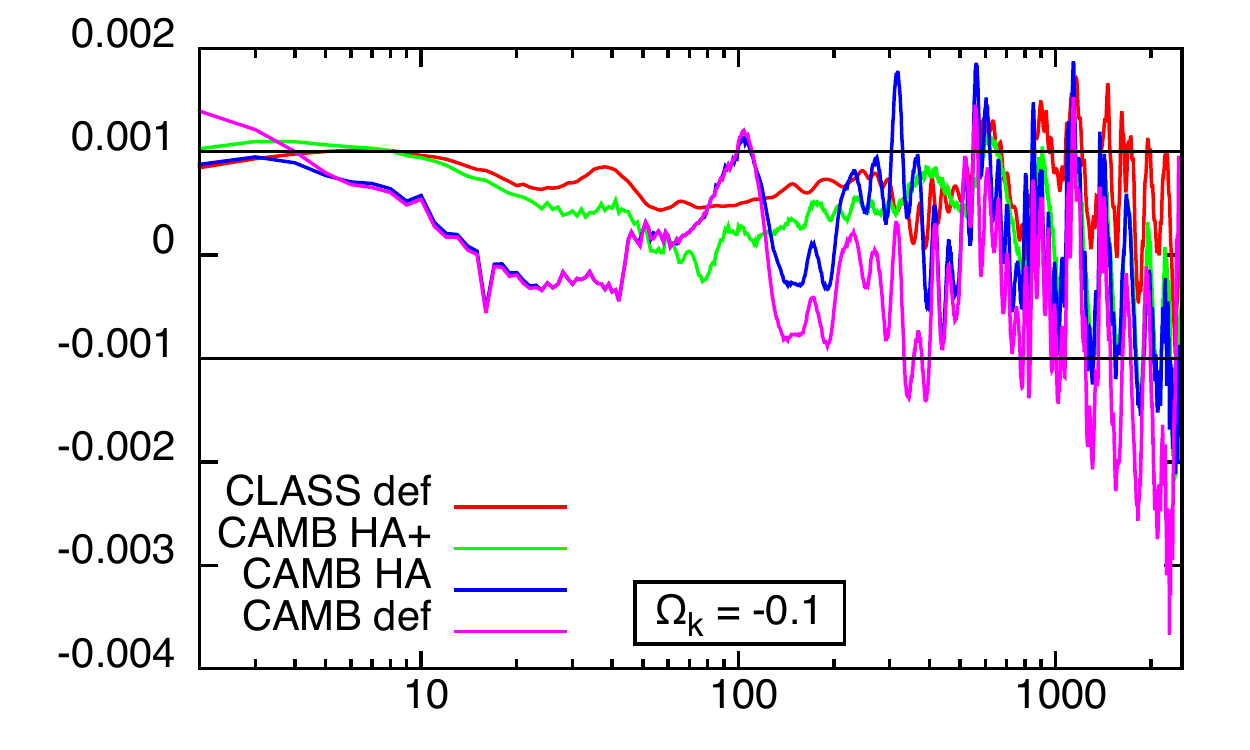}
\includegraphics[width=0.5\columnwidth]{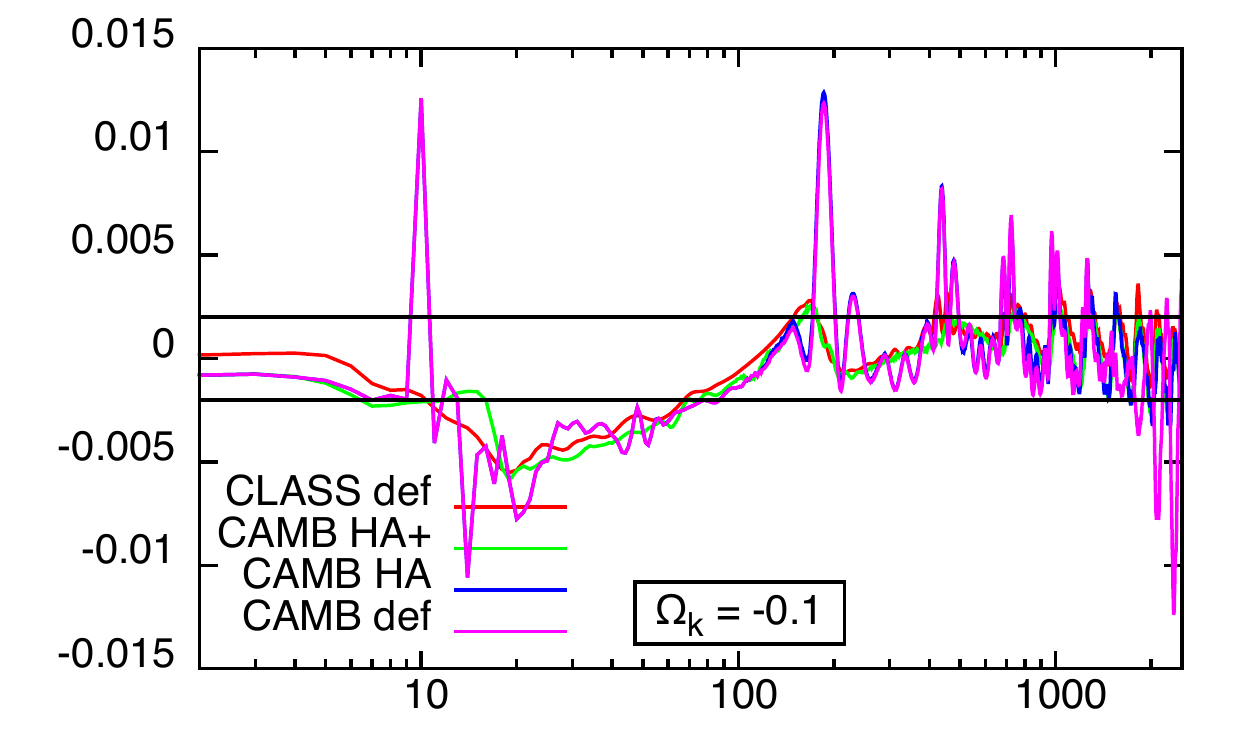}\\
\includegraphics[width=0.5\columnwidth]{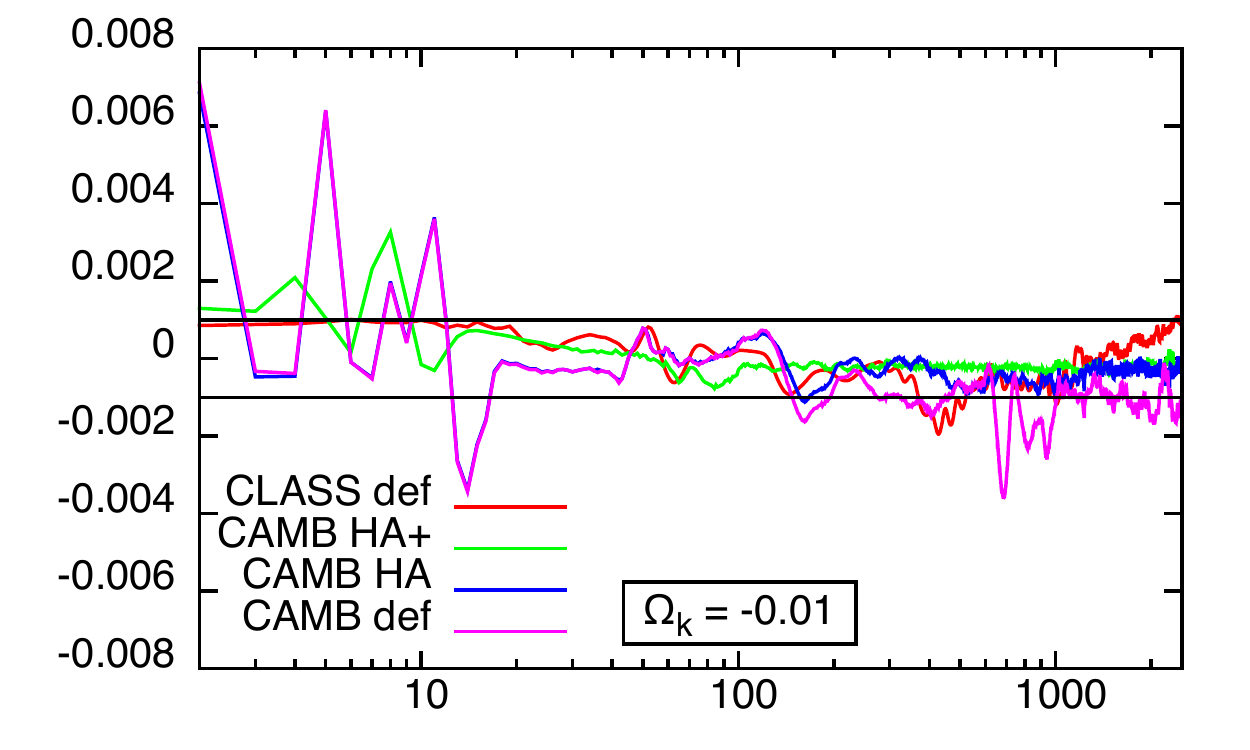}
\includegraphics[width=0.5\columnwidth]{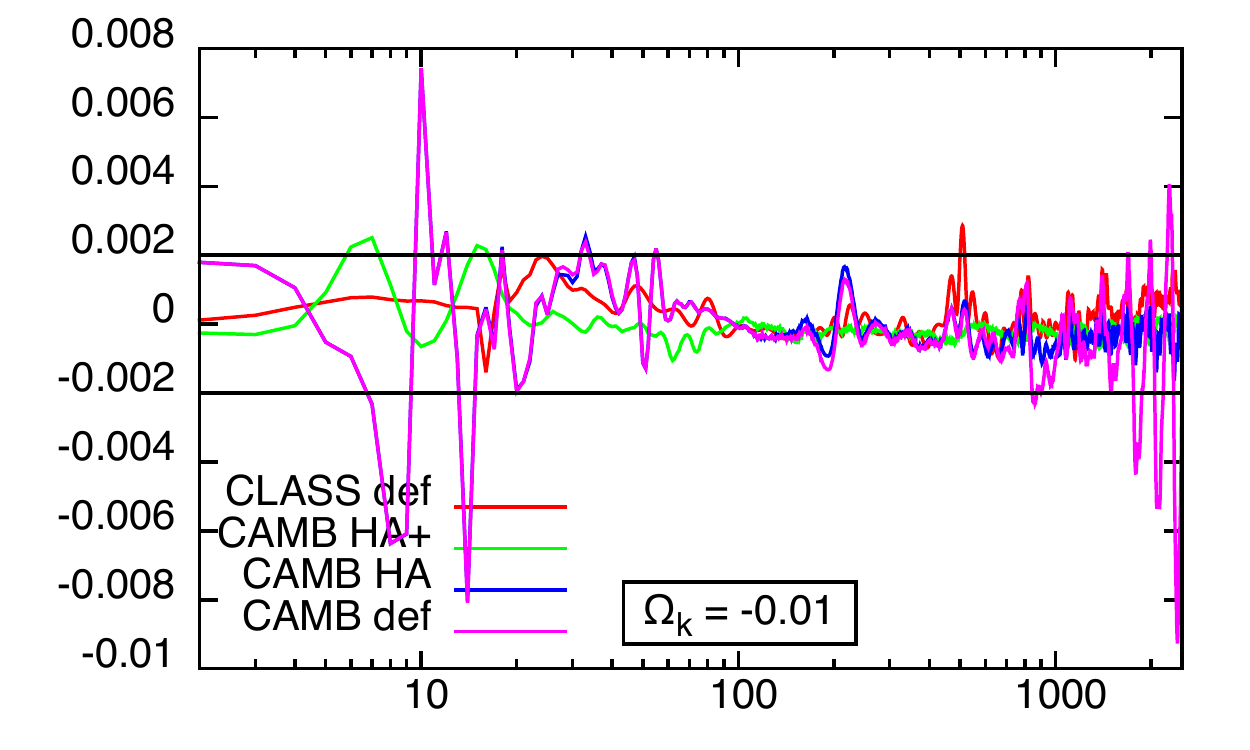}\\
\includegraphics[width=0.5\columnwidth]{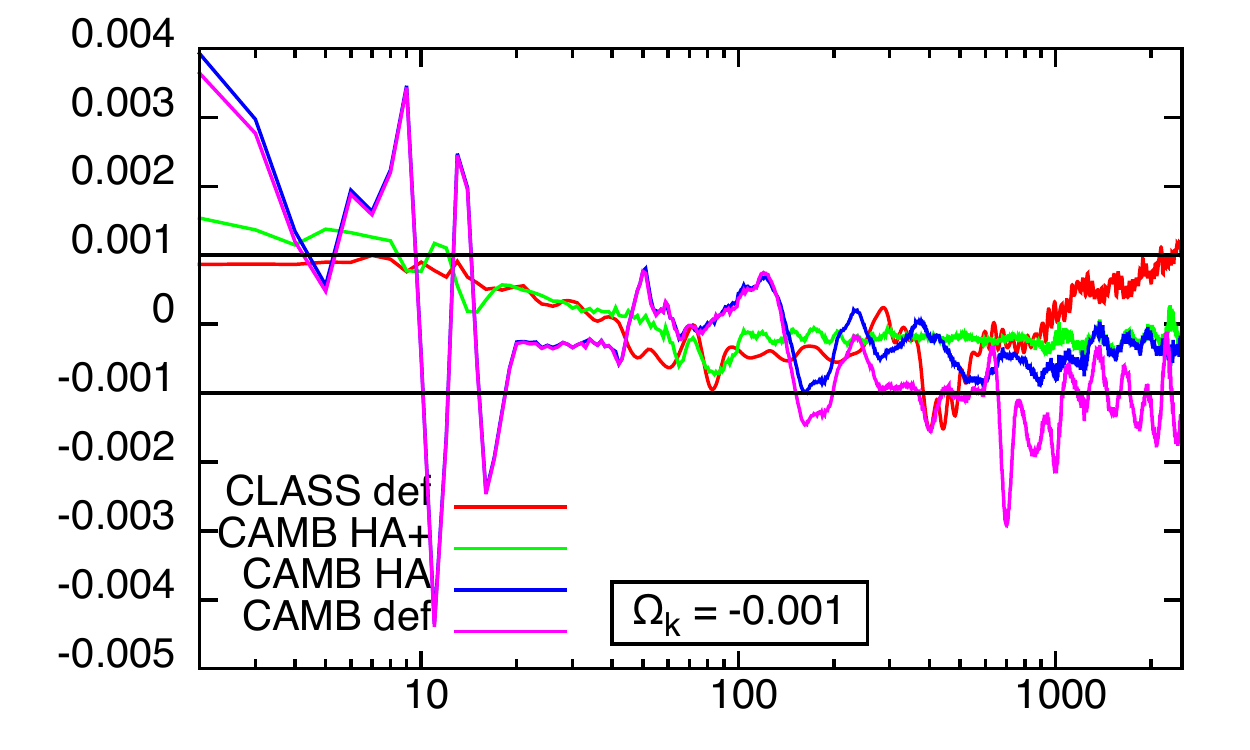}
\includegraphics[width=0.5\columnwidth]{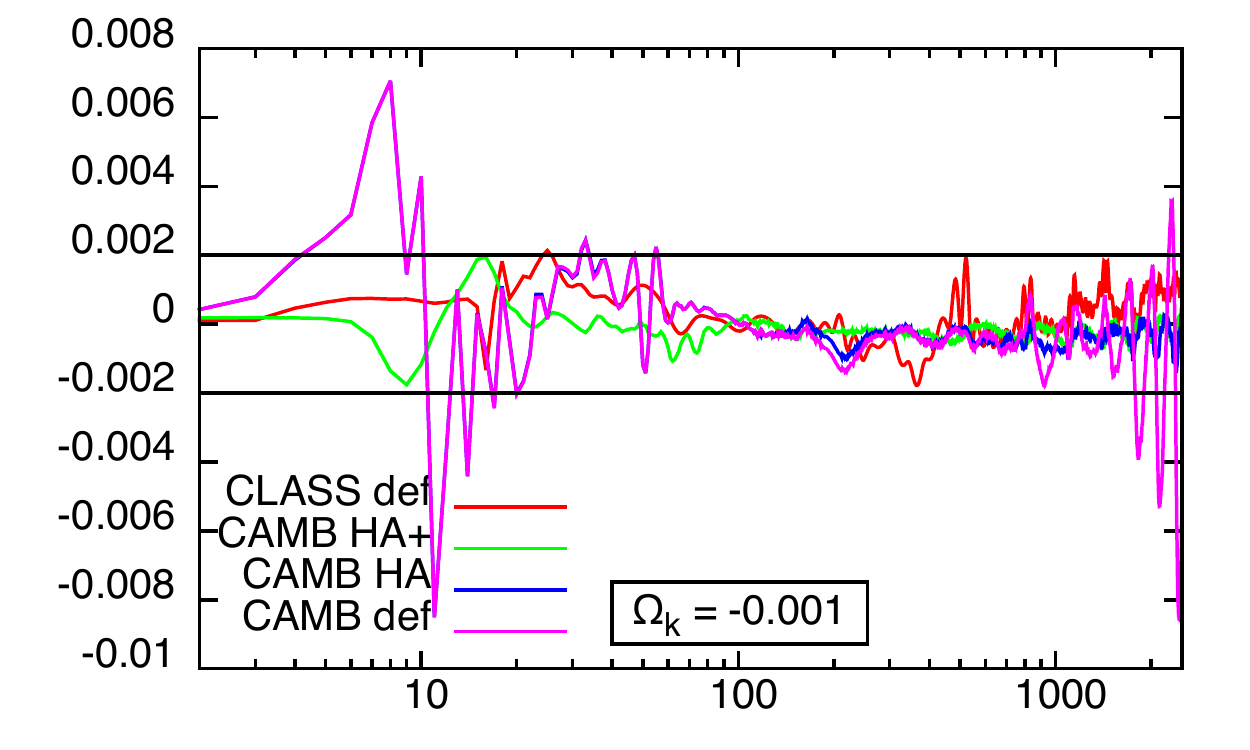}
\caption{Accuracy of \CLASS{} and \CAMB{} in closed $\Lambda$CDM models with, from top to bottom, $\Omega_k=-0.1, -0.01, -0.001$. All spectra for unlensed temperature (left) and E-type polarisation (right) are compared to the \CLASS{} reference settings, argued to be accurate at the 0.01\% level (see text). For \CLASS{}, we show the default settings of v2.0. For \CAMB{}, we show three precision settings: default, ``high accuracy'' (HA), and ``high accuracy'' plus accuracy boost parameters set to 2 (HA+).}%
\end{figure}
\begin{figure}\label{fig:accuracy_open}%
\includegraphics[width=0.5\columnwidth]{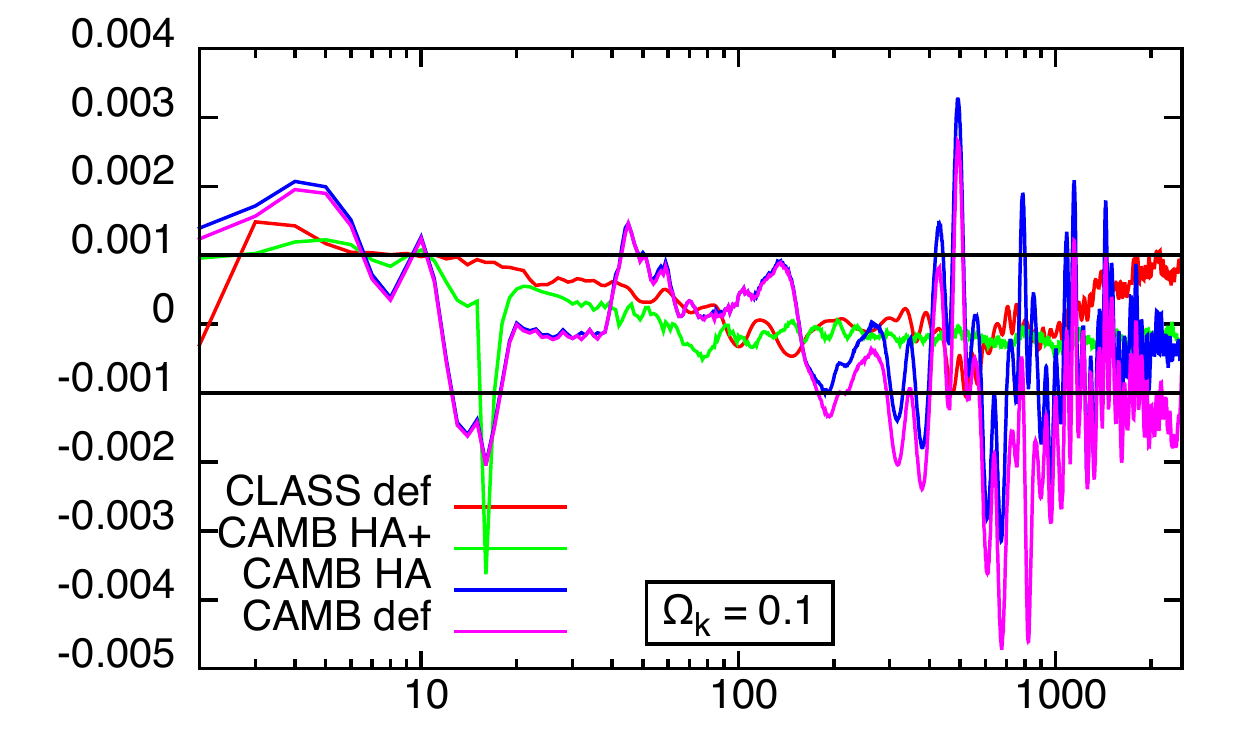}
\includegraphics[width=0.5\columnwidth]{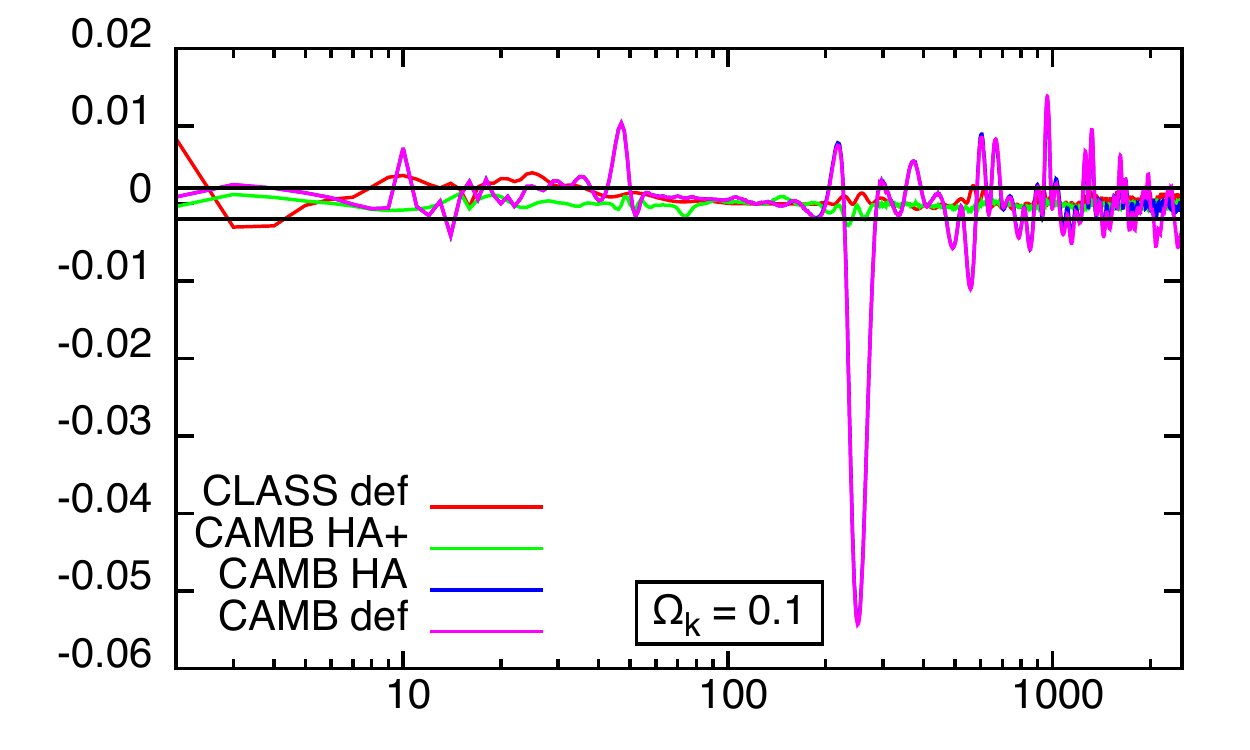}\\
\includegraphics[width=0.5\columnwidth]{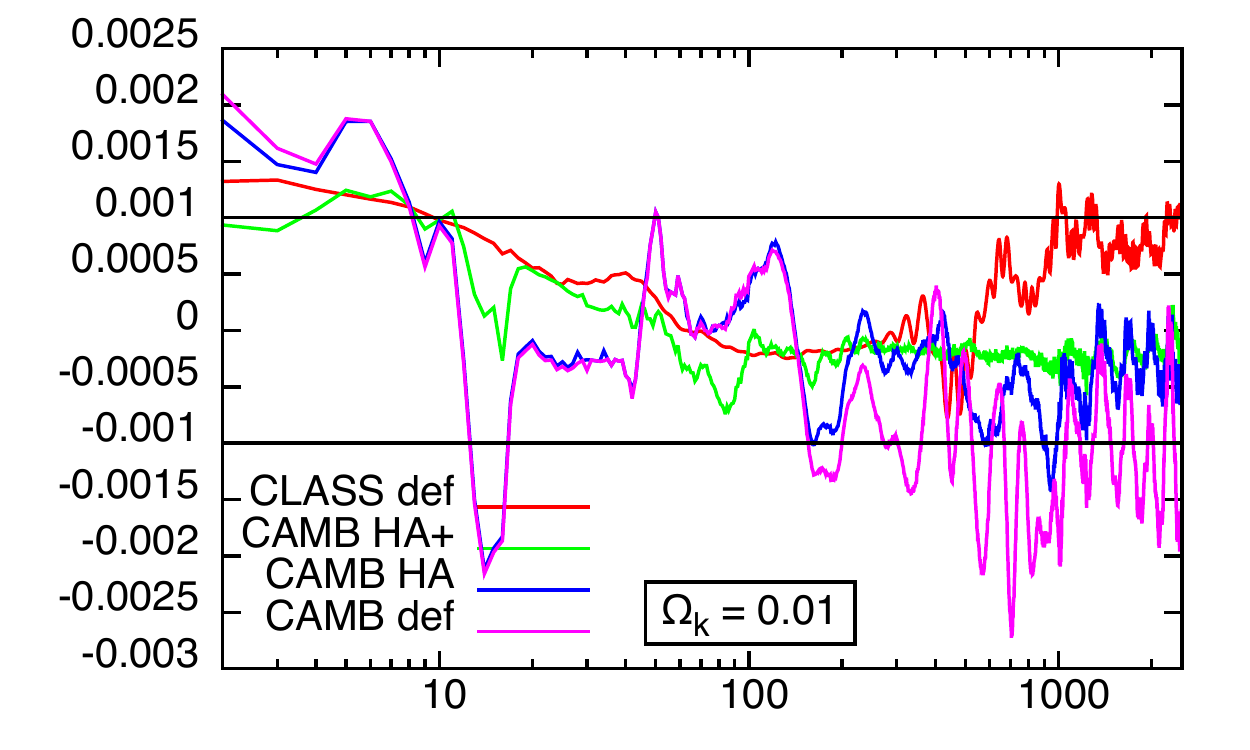}
\includegraphics[width=0.5\columnwidth]{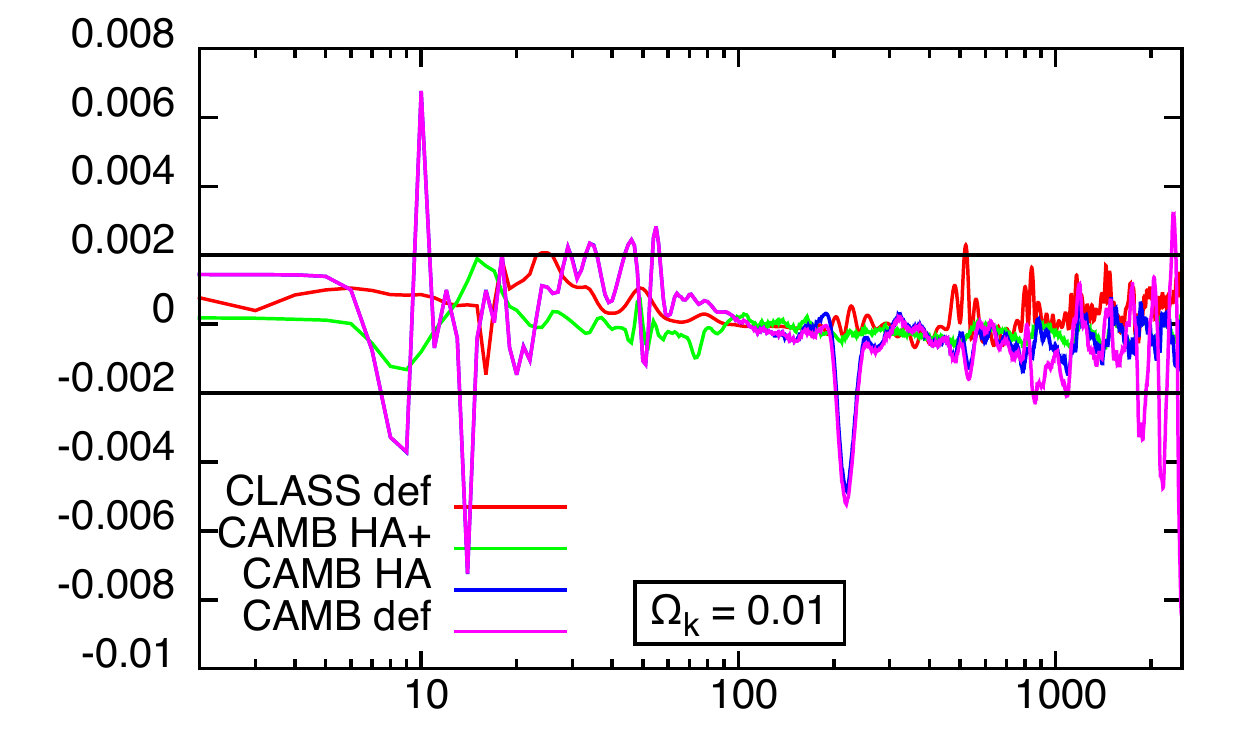}\\
\includegraphics[width=0.5\columnwidth]{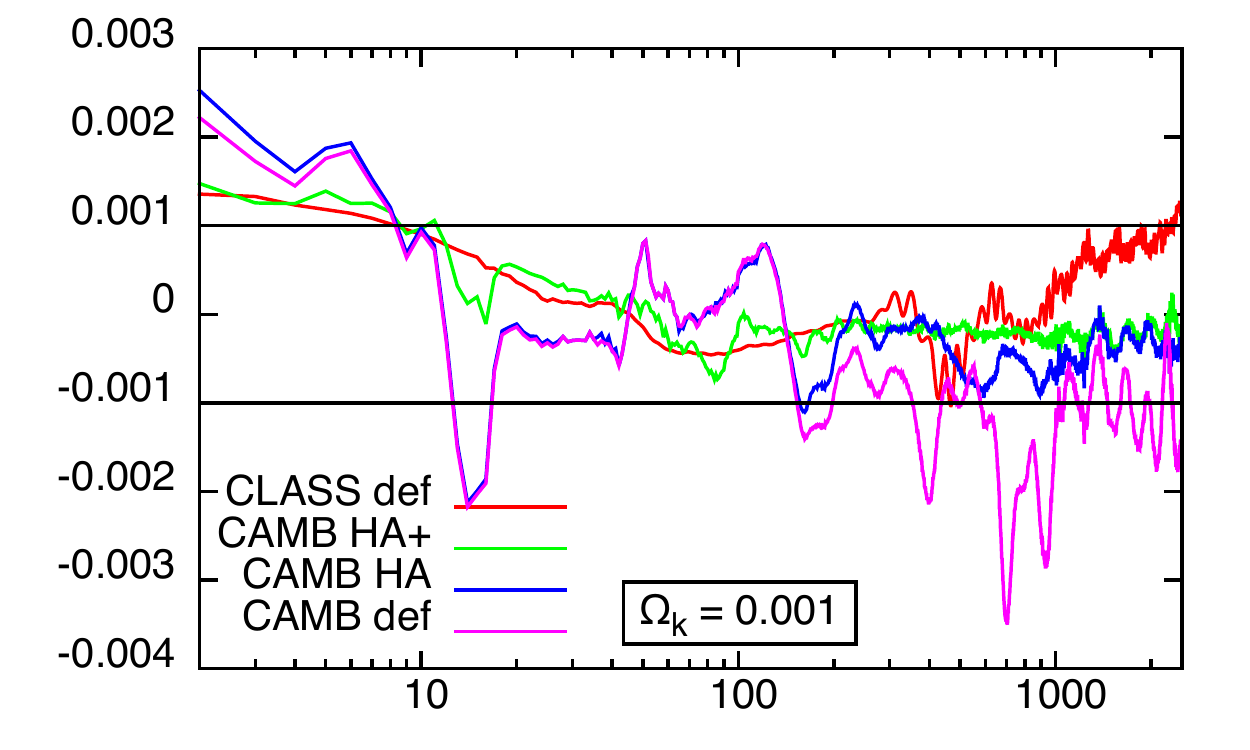}
\includegraphics[width=0.5\columnwidth]{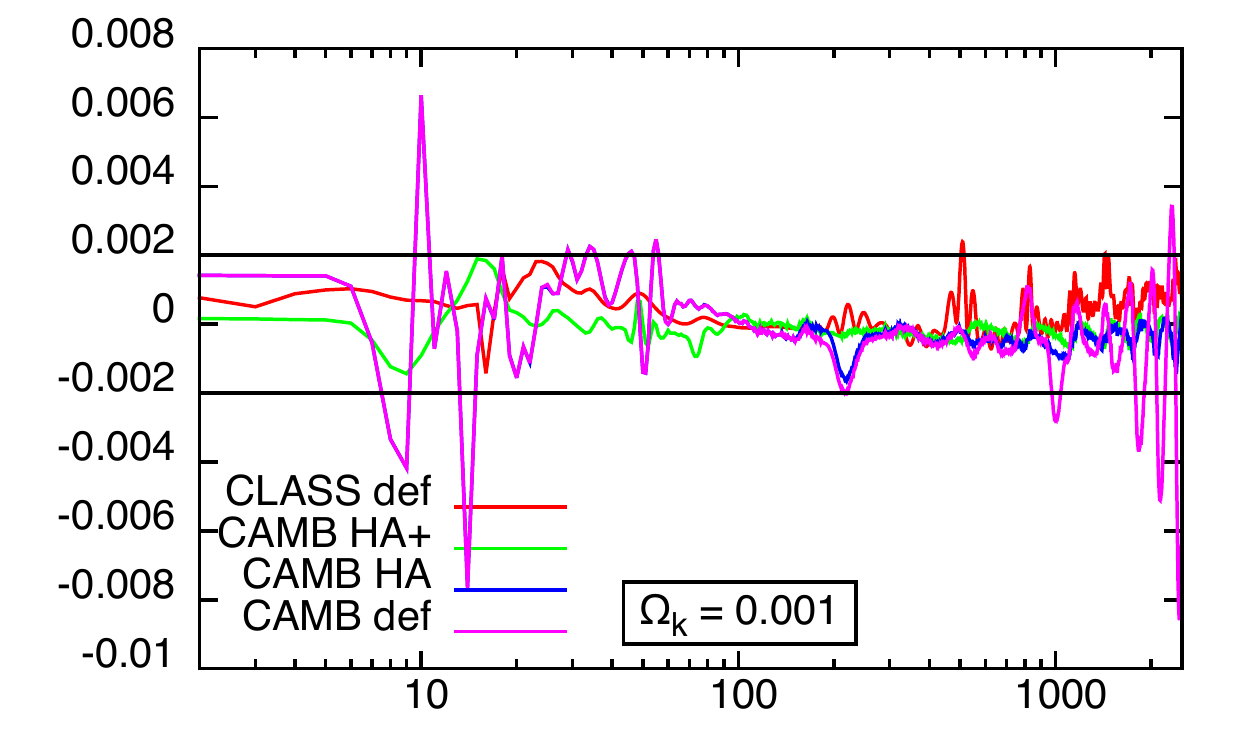}
\caption{Accuracy of \CLASS{} and \CAMB{} in open $\Lambda$CDM models with, from top to bottom, $\Omega_k=0.1, 0.01, 0.001$. All spectra for unlensed temperature (left) and E-type polarisation (right) are compared to the \CLASS{} reference settings, argued to be accurate at the 0.01\% level (see text). For \CLASS{}, we show the default settings of v2.0. For \CAMB{}, we show three precision settings: default, ``high accuracy'' (HA), and ``high accuracy'' plus accuracy boost parameters set to 2 (HA+).}%
\end{figure}

\subsection{Accuracy of \CLASS{} and \CAMB{} in curved $\Lambda$CDM models}

For several values of $\Omega_k$ in the range $\[-0.1:0.1\]$, we tuned the parameters governing the computation of hyperpsherical Bessel function and the sampling of sources and transfer functions in such a way that, like in the flat case, temperature and polarisation scalar spectra converge at the $10^{-3}$\% level (we also checked the convergence of tensor spectra, but at a lower level, given the limited prospects for high-accuracy tensor observations in the future). The reference accuracy settings of the file {\tt cl\_ref.pre} guarantee this level of convergence for $\ell \leq 2500$. In section~\ref{sec:precision flat}, we explained why we believe that in the flat case, this leads to robust ``reference spectra'', precise at the 0.01\% level for scalar temperature. In non-flat models, we expect that reference settings obtained with the same method lead to the same precision. This might not be the case if the code was switching between different algorithms or sampling methods, depending on the value of curvature. The only place in the code where this could be true would be the calculation of Bessel functions, since the starting value for the recurrence method in the closed case is sometimes based on equation~\eqref{eq:Gegenbauer_identity}. However, in the $K\rightarrow 0$, our hyperspherical Bessel algorithm is the same for closed and open models. 

In the rest of the code, no difference is made between the flat and non-flat case: all physical equations are continuous in the variable $K$, we use the same routines and the same sampling strategies everywhere for flat and non-flat models, and we switch off the flat rescaling approximation, which is specific to non-flat models, when we establish reference settings. Hence, we expect our non-flat reference setting to provide the same accuracy as in the flat case, namely 0.01\% for scalar temperature and 0.02\% for scalar polarisation.

Like in the flat case, we degraded the precision parameters in such a way that, by default, the code achieves roughly 0.1\% precision on scalar temperature spectra, or 0.2\% on scalar polarisation, throughout the range $-0.1<\Omega_k<0.1$ and for $\ell \leq 2500$. The resulting error is shown in figures~\ref{fig:accuracy_closed} and~\ref{fig:accuracy_open} for several values of $\Omega_k$. In the same figures, we show \CAMB{} errors for the same precision choices as before: default, with ``high accuracy'', and finally with further increasing the three ``accuracy boost parameters'' from 1 to 2. The fact that the last setting converges towards the \CLASS{} reference model at the level of 0.03\% (except for $\Omega_k=-0.1$) is important: it brings the first independent test of the fact that the method implemented in \CAMB{} for hyperspherical Bessel functions can reach higher accuracy than what is requested by current and future CMB data.

On the other hand, the default precision settings in \CAMB{} turn out to produce rather large errors in the large curvature limit: up to 0.5\% for temperature and 1\% for polarisation. With ``high accuracy'' settings, the low multipoles remain equally inaccurate, but in the region where high precision is really needed, i.e. for $500 < \ell < 2500$, the ``high accuracy'' settings lead to roughly 0.1\% accuracy on temperature for all values of $\Omega_k$ with the exception of $\Omega_k=-0.1$. We conclude that with the current versions of the two codes, \CAMB{} with ``high accuracy'' is nearly as precise as \CLASS{} with default settings, even for non-flat models. For these settings, a comparison of running time (see in Table~\ref{tab:performance2}) shows that \CLASS{} is typically 2 to 3 times faster than \CAMB{} for non-flat models. This can be attributed to the efficient algorithmic implementation described in section~\ref{sec:hyper} as well as the flat rescaling approximation. 
\begin{table}\label{tab:performance2}
\begin{center}
\begin{tabular}{|c|ccccccc|}
\hline
$\Omega_k$   & 0     & $10^{-3}$ & $10^{-2}$ & $10^{-1}$ & $-10^{-3}$ & $-10^{-2}$ & $-10^{-1}$ \\
\hline
\CLASS{} & 9.7 & 29.7 & 21.7 & 43.4 & 24.5 & 27.7 & 79.4 \\
\CAMB{}  & 6.4 & 94.7 & 89.9 & 87.2 & 68.4 & 72.3 & 163.7 \\
\hline
\end{tabular}
\end{center}
\caption{Execution time of \CLASS{} (default precision) and \CAMB{} (with ``high accuracy'' settings) for different values of $\Omega_k$. In these settings, \CLASS{} is slightly more accurate, especially for low $\ell$'s and large positive curvature. We asked the codes to compute the lensed temperature and polarisation scalar spectra up to the same $l_\mathrm{max}=3000$. For a reliable comparison, we switched off {\tt OpenMP},  and compiled the codes with recent versions of {\tt gcc} or {\tt gfortran}, using exactly the same optimisation flags, and running on the same processor. These times are expressed in seconds, but they are only useful for the sake of comparison: on modern processors with many cores the execution time will of course be much smaller.}
\end{table}

\subsection{Continuity across $\Omega_k=0$}

\begin{figure}\label{fig:accuracy_continuity}%
\includegraphics[width=0.5\columnwidth]{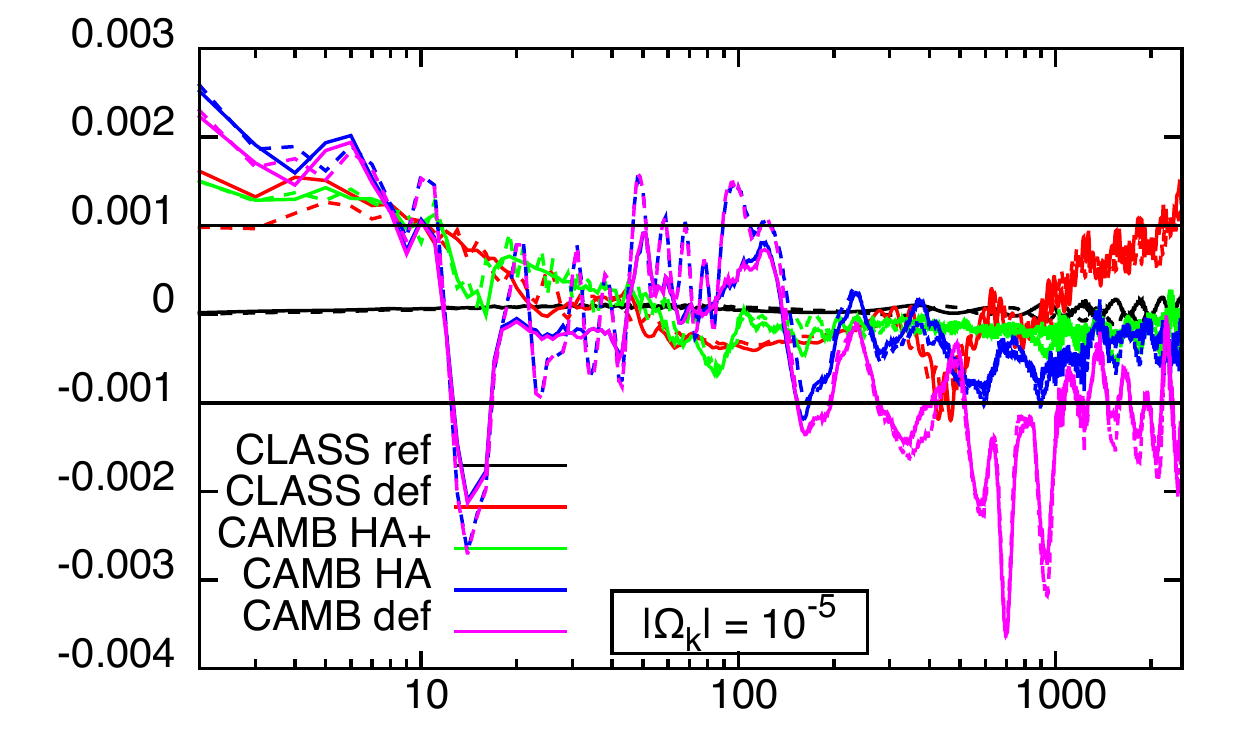}%
\includegraphics[width=0.5\columnwidth]{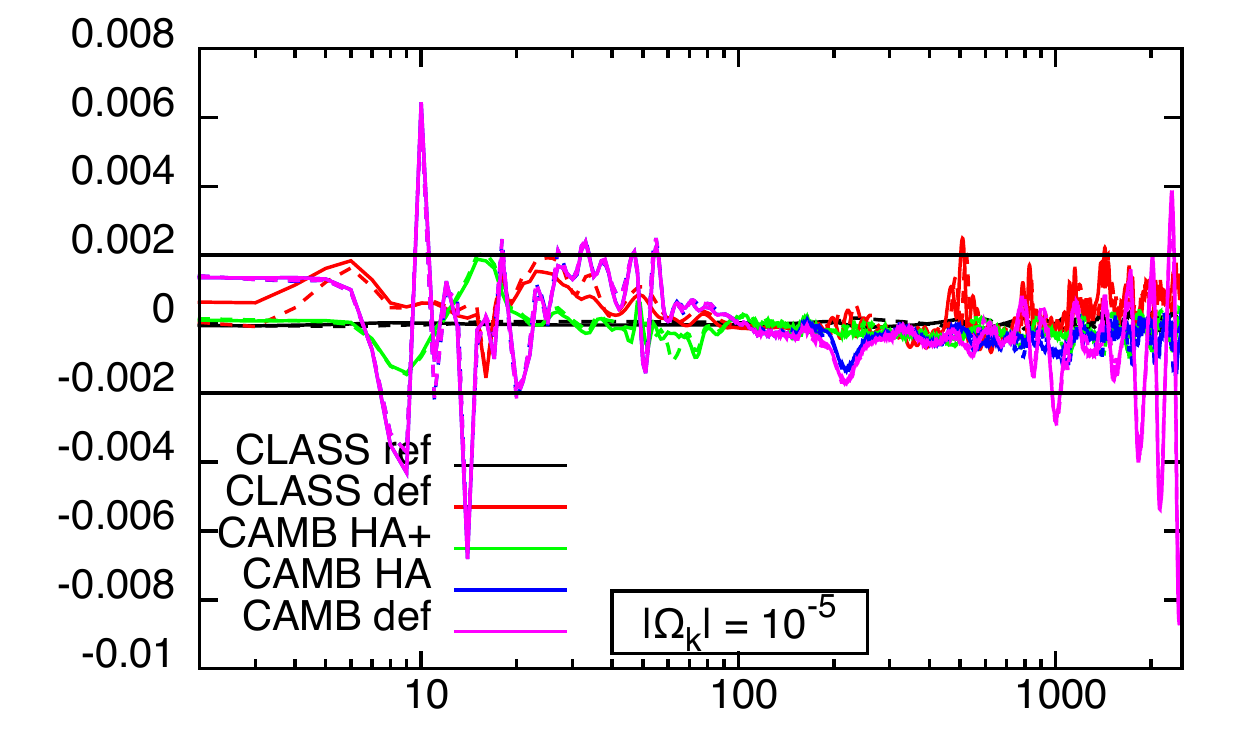}%
\caption{Test of the continuity of the spectra across $\Omega_k=0$. The results for $\Omega_k=10^{-5}$ (solid curves) and $\Omega_k=-10^{-5}$ (dashed curves) are compared to the 
\CLASS{} reference \emph{flat} model, for unlensed temperature (\emph{left panel}) and E-type polarisation (\emph{right panel}). For  \CLASS{}, we show the reference settings and default settings of v2.0. For \CAMB{}, we show three precision settings: default, ``high accuracy'' (HA), and ``high accuracy'' plus accuracy boost parameters set to 2 (HA+).}%
\end{figure}

Checking the continuity of the $C_\ell$'s across the special value $\Omega_k=0$ can be seen as a further test of precision. One might fear that, when all cosmological parameters but $\Omega_k$ are fixed, a discontinuity could be observed in $\Omega_k=0$, either due to insufficient accuracy settings or a real issue with the algorithms used in the code (for instance, in the calculation of closed or open hyperspherical Bessel functions, or due to the fact that $q$ takes discrete values in a closed universe or arbitrary values in an open one). In a Monte Carlo run for parameter extraction, such a discontinuity could cause a significant problem and bias the final results.

In figure~\ref{fig:accuracy_continuity}, we show the ratio of temperature and polarisation spectra computed for $\Omega_k=\pm10^{-5}$ compared to the \CLASS{} reference  spectra for $\Omega_k=0$. We expect the true difference between the spectra to be also of the order $10^{-5}$. The precision settings to be used in a parameter extraction code should be such that these models do not differ by more than 0.1\%, meaning that they could not be discriminated from each other, given the accuracy of current CMB data.

We check that with \CLASS{} and reference precision settings, the spectra indeed agrees at the $10^{-5}$ level. With default settings, \CLASS{} is supposed to produce an error of the order of 0.1\% at most. This is consistent with the difference observed in figure~\ref{fig:accuracy_continuity} between the $\Omega_k=0$ and $\Omega_k=\pm10^{-5}$ spectra. Actually, in this case continuity across $\Omega_k=0$ is achieved automatically due to the flat rescaling approximation. For \CAMB{}, the differences are at the expected level: 0.3\% for temperature with default settings, 0.1\% with ``high accuracy'', and even less with all boost parameters set to 2. We conclude that the two codes successfully deal with the  $\Omega_k \rightarrow 0$ limit, and do not suffer from artificial step effects between closed and open models. 

\subsection{Lensing spectrum and lensed CMB spectra}

The previous sections refer to the accuracy of unlensed temperature/polarisation spectra. The lensed spectra are obtained as a function of the unlensed ones and of the CMB lensing potential spectrum $C_\ell^{\phi \phi}$. In this section, we will test the accuracy with which this quantity is computed by \CLASS{} in curved space. This is crucial to establish the accuracy of lensed spectra. Instead, the step leading from unlensed to lensed CMB spectra, coded in the {\tt lensing} module of CLASS, is exactly the same in flat and curved space; its accuracy has been extensively tested in \cite{Lesgourgues:2011rg}, and there is no need to present new tests in curved space.

The CMB lensing spectrum $C_\ell^{\phi \phi}$ follows from the CMB lensing transfer function ${\Delta_\ell^\phi}(q,\tau_0)$ through Eq.~(\ref{eq:spec1}). For low $\ell$'s, this transfer function is computed by integrating over time,
\begin{equation}
{\Delta_\ell^\phi}(q,\tau_0) = \int_{\tau_\mathrm{rec}}^{\tau_0} d\tau S_{\phi}(k,\tau) \Phi^{\nu}_{\ell}(\chi)~, \label{eq:delta_phi}
\end{equation}
where the relations between $k$, $q$, $\nu$, $\chi$ are the same as in the rest of this paper. This integral is of the same form as the one leading to the first temperature transfer function ${\Delta_\ell^{T_0}}(q,\tau_0)$, except for the shape of the source function: for CMB lensing, the source is very smooth and very broad in the range $\tau_\mathrm{rec} < \tau < {\tau_0}$. Previous tests presented in this section show that the \CLASS{} default precision settings 
%($\tau$ step, $q$ steps, accuracy of the hyperspherical bessel functions $\Phi^{\nu}_{\ell}(\chi)$) 
are sufficient for getting an accurate and well-sampled temperature transfer function. Nevertheless, given the very different shape of the CMB lensing source function, similar tests should be repeated for the lensing potential.

The smoothness of $S_{\phi}(k,\tau)$ allows to use the Limber approximation for large $\ell$'s, i.e. to replace the integral by a simple evaluation of the source function at a given time. Details on this approximation (in flat and curved space) are discussed in Appendix~\ref{sec:limber}.

\begin{figure}\label{fig:accuracy_lensing}%
\includegraphics[width=0.5\columnwidth]{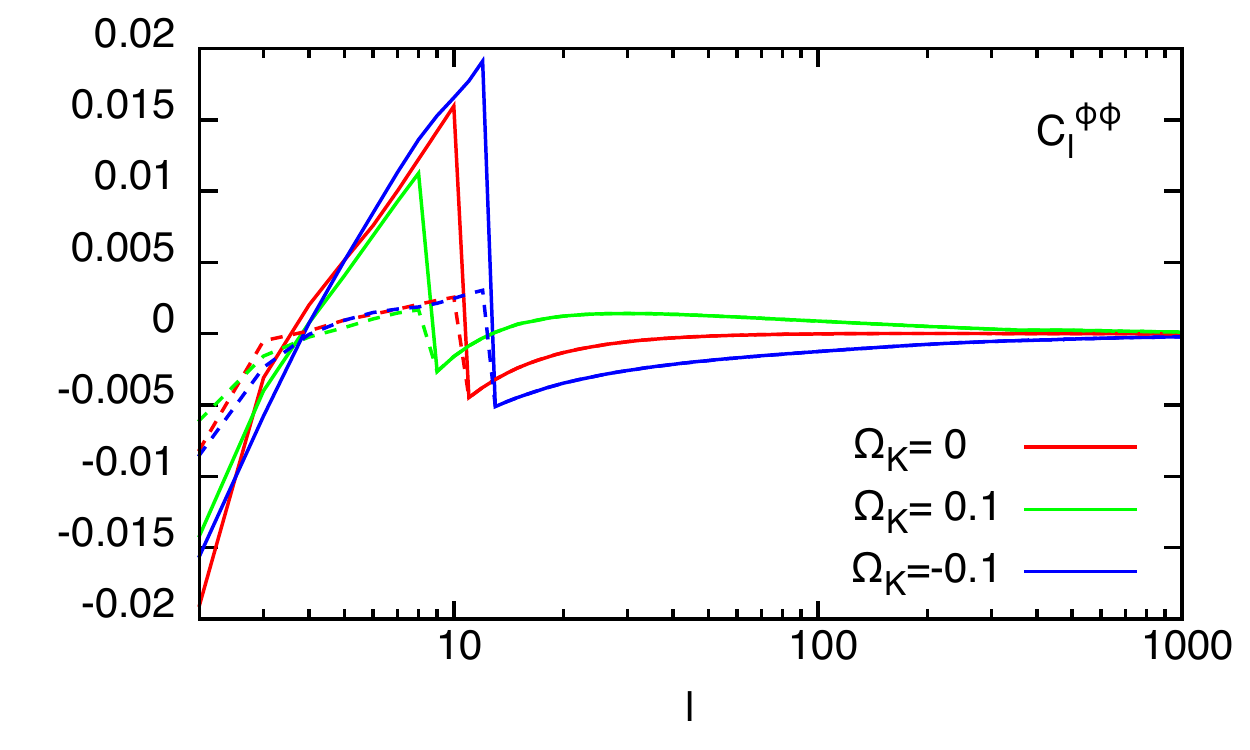}%
\includegraphics[width=0.5\columnwidth]{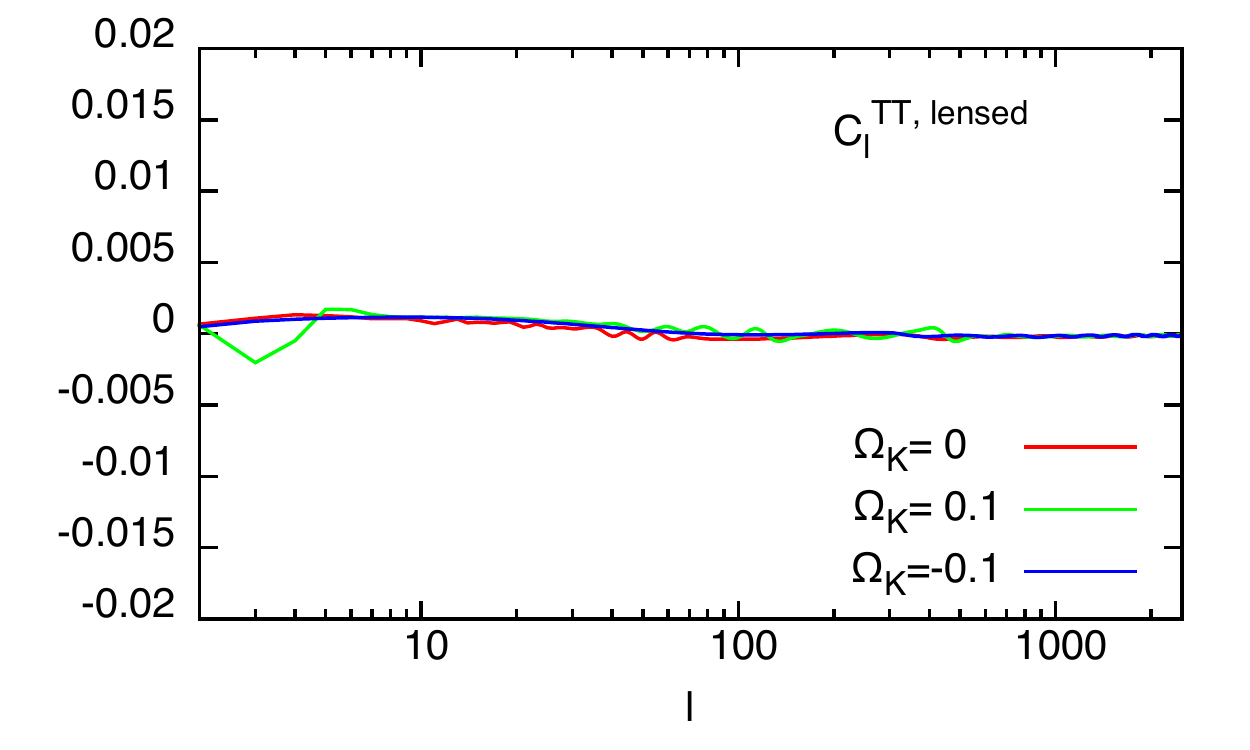}%
\caption{Test of the accuracy of the CMB lensing potential $C_\ell^{\phi \phi}$ (left plot), and of the propagation of the error in $C_\ell^{\phi \phi}$ to the lensed power spectra (right plot). For $\Omega_k=0$, $0.1$, $-0.1$, we ran \CLASS{} with reference settings for the parameters governing the accuracy of  $C_\ell^{\phi \phi}$  (in particular, not using the Limber approximation), and then with default precision. We show the ratio of CMB lensing potential spectra on the left, and the corresponding ratio of lensed temperature spectra on the right. At low-$\ell$, before using the Limber approximation, the default $C_\ell^{\phi \phi}$ are not precise, but the impact of this error on lensed spectra is negligible; precise $C_\ell^{\phi \phi}$ at low $\ell$ can be obtained by just halfing the precision parameter {\tt source\_sampling\_step} (leading to the dashed curves). These tests were performed later than the rest of this paper, with \CLASS{} v2.3.3.}%
\end{figure}

To check the accuracy of $C_\ell^{\phi \phi}$ in the range which is most relevant for CMB lensing, we computed some reference spectra for various values of $\Omega_k$, ensuring that:
\begin{itemize}
\item the Limber approximation was never used in the range $2 \leq \ell \leq 1000$,
\item all precision parameters governing the $\tau$-sampling in the integral, the $q$-sampling of transfer functions, and the accuracy with which hyperspherical bessel functions $\Phi^{\nu}_{\ell}(\chi)$ are computed were pushed to extreme values such that the $C_\ell^{\phi \phi}$'s are fully converged and stable (up to the level of $10^{-5}$).
\end{itemize}
Then, we computed the same $C_\ell^{\phi \phi}$ with default accuracy settings (i.e. with poor precision settings for the integral at low $\ell$, and using the Limber approximation for $\ell \geq {\tt l\_switch\_limber}=10$). We show the ratio of the default over reference spectra in figure~\ref{fig:accuracy_lensing} (left plot). 

For $l<10$, the error is large, up to 2\%. The settings leading to accurate small-scale temperature spectra do not lead to accurate CMB lensing spectra. This can easily be fixed by just increasing the time sampling in the integral~(\ref{eq:delta_phi}): by halfing the precision parameter {\tt source\_sampling\_step}, we reach 0.5\% precision (dashed curves in figure~\ref{fig:accuracy_lensing}). However, there is no need to change the default precision settings just for improving low-$l$ $C_\ell^{\phi \phi}$'s, because on very large angular scales the lensing spectrum is impossible to measure accurately, and impacts the final lensed $C_\ell$'s by a negligible amount, as we shall see in the next paragraph. For $l\geq 10$, we see the power of the Limber approximation: it induces an error of at most $0.5\%$ near $l=10$, and it becomes increasingly good at large $\ell$'s. 

To check how errors in the lensing spectrum propagate to lensed $C_\ell$'s, we computed the ratio of lensed temperature spectra for the same pair of models. We find an accuracy level better than 0.1\% (except for $l=3$ in the open model), clearly sufficient for fitting current and forthcoming CMB experimental data. Hence, the 2\% error on the very low $\ell$ CMB lensing spectrum has a negligible impact on lensed temperature spectra, which are mainly sensitive to larger $\ell$'s. For default accuracy, we found in figures \ref{fig:accuracy_flat}, \ref{fig:accuracy_closed} , \ref{fig:accuracy_open} that the error in the unlensed spectra is of the order of 0.2\%; figure \ref{fig:accuracy_lensing} shows that in the lensing spectra the error may increase at most to the level of 0.3\%.

\section{Discussion}

Having efficient Boltzmann codes for non-flat models is still useful, despite the fact that the data is compatible with spatial flatness. Indeed we will have 
soon more accurate Planck temperature data, Planck polarisation data and accurate polarisation data coming from other ground-based experiments. In addition, there will be new generations of large scale structure experiments in the future and maybe also another CMB experiment. Each time that new data arrives, we will need to check that the flat model is still preferred, and to derive bounds on curved models. 

Previous implementations of curvature in Boltzmann codes were not thoroughly tested due to the lack of independent methods. Moreover, the method encoded in \CAMB{} for computing hyperspherical Bessel function is difficult to test because it is built into the computation of the integrals. Here we have presented a new method for calculating CMB anisotropies in a non-flat FLRW universe, relying on a very stable algorithm for the calculation of hyperspherical Bessel functions, that can be pushed to arbitrary precision levels.  We also introduce a new approximation scheme which gradually takes over in the flat space limit, and significant speeds up calculations. We described several aspects of our implementation of the equations in the code (sources, radial functions, transfer functions, primordial spectra) aimed at simplicity and unification of flat and non-flat computations.

We used our code to  benchmark the accuracy of the \CAMB{} code in curved space: by default \CAMB{} achieves roughly 0.3\% on scalar temperature, or 0.1\% with the ``high accuracy'' flag tuned on. This is roughly what was claimed before, and it is similar to what one gets in the flat case. The exception is the limit of large positive curvature, since our comparison reveals slightly larger errors for $\Omega_k=-0.1$. We will check in the future that we obtain the same lower bound on $\Omega_k$ when using \CLASS{} instead of CAMB{} to analyse Planck data, but this is likely to be the case up to insignificant differences.

However we find that for the same precision level, \CLASS{} is significantly faster, usually by a factor 3 for non-flat models. This is mainly due to the flat rescaling approximation, which could in principle be easily implemented in other codes.

\section*{Acknowledgements} This project was supported by the Swiss National Foundation.

\appendix

\section*{Appendix}
The appendix describe 4 approximation schemes used by \CLASS{}. The splitting of the temperature source function described in appendix~\ref{app:sources} is unique to~\CLASS{}, but the time cut approximation and the multipole cut approximation of appendix \ref{sec:time_cut} and \ref{sec:multipole_cut} may have been used in previous Boltzmann codes in some form. (They are rather straightforward.) However, this is the first time the physical motivations for these approximations have been published. The Limber approximation of appendix \ref{sec:limber} is more standard.

\section{Splitting the temperature source functions \label{app:sources}}

There are several ways to split the temperature source functions in a set of physical contributions: this is just a matter of convention. Here we refer to the most common splitting. To make it more readable, we reorganize the contributions to  $S_{T_0}$, $S_{T_1}$, $S_{T_2}$ in a way which differs from both (\ref{eq:sourceT}) and (\ref{eq:sourceTbis}), but is still fully equivalent after some integrations by part:  
\begin{equation}
S_{T0}^{(0)}= g \left( \frac{1}{4}\delta_\gamma + \psi \right) + e^{-\kappa} (\phi^\prime+\psi^\prime)~,~~~~
S_{T1}^{(0)}= \frac{g}{k} \theta_b ~,~~~~
S_{T2}^{(0)}=g P^{(0)}~. \label{eq:sourceTter}
\end{equation}
The first term  in $S_{T0}^{(0)}$ contains the intrinsic temperature fluctuation ($\frac{1}{4}\delta_\gamma$) and the gravitational redshift term $\psi$. The words ``Sachs-Wolfe term" sometimes refer to the latter, or to the sum of the two: here we call the whole term proportional to $g$ ``Sachs-Wolfe". The second term proportional to $e^{-\kappa}$ is the Integrated Sachs-Wolfe term. $S_{T1}^{(0)}$ contains the Doppler term, and $S_{T2}^{(0)}$ conatins some polarisation-related contributions (that would vanish if the Thomson scattering term was averaged over directions). We can now write the source terms (\ref{eq:sourceTbis}) implemented in \CLASS{} with a set of switching coefficients $\{ s_\mathrm{SW}, s_\mathrm{ISW}, s_\mathrm{Dop.}, s_\mathrm{Pol.}\}$ which should all be set to 1 in order to recover the full temperature spectrum, or some of them can be set to 0 in order to kill each of the four physical contributions:
\begin{eqnarray}
\tilde{S}_{T_0}^{(0)} &=& s_\mathrm{SW} \left[ g\left( \frac{1}{4}\delta_\gamma+\psi \right)\right] + s_\mathrm{ISW} \left[ g(\phi-\psi) +e^{-\kappa} 2 \phi' \right] 
+ s_\mathrm{Dop.} \left[ k^{-2}(g \theta_b'+g' \theta_b)\right]~, \nonumber\\
\tilde{S}_{T_1}^{(0)} &=& s_\mathrm{ISW} \left[ e^{-\kappa} k (\psi-\phi)\right]~,~~~~
\tilde{S}_{T_2}^{(0)} = s_\mathrm{Pol.} \left[ g P^{(0)} \right]~.
\end{eqnarray}
In version 2.0 of \CLASS{}, we decomposed the expression (\ref{eq:sourceTbis}) of the source functions in this way, to allow users to switch off some terms when necessary (e.g, to study ISW correlations with large scale stucture, or to understand the physically impact of some cosmological ingredient). We have shown this feature in the right panel of figure~\ref{fig:decomposition}. As explained in the file {\tt explanatory.ini}, the \CLASS{} user can specify a list of terms to be included in the temperature calculation. In this list, the ISW terms has been further split into early and late ISW\footnote{For instance, to include only the late ISW term, the user should write {\tt temperature contributions = lisw} in the input file. For all contributions except the late ISW, the syntax would be {\tt temperature contributions = tsw, eisw, dop,pol}. The splitting between early and late ISW occurs at an arbitrary redshift which can be adjusted by passing a value for the input parameter {\tt early/late isw redshift} (set by default to 120).}

\section{Time cut approximation \label{sec:time_cut}}

The lower boundary of the conformal time integral in equations~(\ref{eq:transT}--\ref{eq:transB}) is found automatically by \CLASS{}: it is the time at which the Thomson scattering rate $\kappa'=an_e\sigma_T$ exceeds a given fraction of the Hubble rate\footnote{Defining the characteristic times $\tau_c=1/\kappa^\prime$ and $\tau_H=a/a^\prime$, the lower boundary is the time at which $\tau_c/\tau_H = 0.008$ (default setting), or $\tau_c/\tau_H = 0.006$ (high precision setting).}. The upper boundary is by default $\tau_0$, the time today. However it gets automatically reduced by the code for large $\ell$'s, using the fact that for large $\ell$ and small $\chi$, $\Phi_\ell^\nu(\chi)$ (or $j_l(\nu \chi)$) is negligible. All Bessel functions are approximated by zero when they are below a given threshold $\Phi_*$\footnote{In \CLASS{} v1.x, this value was called {\tt bessel\_j\_cut} and fixed by default to $10^{-5}$. In \CLASS{} v2.0, since the whole calculation of Bessel functions was revised as described in Section~\ref{sec:hyper}, this parameter has changed: it is called {\tt hyper\_phi\_min\_abs} and fixed to $10^{-10}$}, and this translates into an upper bound of the conformal time integral.

In \CLASS{} version v2.0, we introduced an additional approximation, allowing to further reduce the upper bound. The source terms $\tilde{S}_{T_2}^{(0)}$ and $S_P^{(m)}$ are proportional to the visibility function $g$. This function has a large peak near the recombination time, and a much smaller peak near the reionisation time. Small angular scales are not sensitive to the effect of rescattering at reionisation, and feel the imprint of the source terms $\tilde{S}_{T_2}^{(0)}$ and $S_P^{(m)}$ only around the time of recombination. For these scales, we can cut the integral over time soon after recombination, when the visibility function goes below a threshold $g_c$. This approximation should not be used on large angular scales, which are sensitive to the small peak of the visibility function near the reionisation time. In conclusion, the time cut approximation consists in cutting the time integral for $\tilde{S}_{T_2}^{(0)}$ and $S_P^{(m)}$ at a time $\tau_c$ such that $g(\tau_c) = g_c$, but only for $l < \alpha \, l_c$ ($\alpha$ is the angular rescaling factor defined in equation~(\ref{eq:alpha}), equal to one in flat space). In the default version of \CLASS{}, these parameters\footnote{Inside the code, $g_c$ and $l_c$ are called respectively {\tt neglect\_CMB\_sources\_below\_visibility} and {\tt transfer\_neglect\_late\_source}.} are fixed to $(g_c, l_c)=(10^{-3}, 400)$, while in the reference settings of the file {\tt cl\_ref.pre} they are pushed to $(10^{-30}, 3000)$, such that the approximation is never used.

In the case of $\tilde{S}_{T_1}^{(0)}$, we can use the same approximation in order to save extra time, because this term accounts for one part of the late ISW effect, and only contributes to small multipoles. With default settings, the time cut does not alter any calculation for $l<400 \alpha$. This covers the whole region in which the ${T_1}$ term contributes to the total ${C_\ell^{TT}}^{(0)}$'s at more than 0.01\%, and to the late ISW ${C_\ell^{TT}}^{(0)}$'s at more than one per cent\footnote{Moreover, on small angular scales with $l>400 \alpha$, the late ISW signal is very difficult to observe, even when cross-correlating temperature and large scale structure maps.}. Hence  the time cut approximation can be safely used for $T_1$.

\begin{figure}\label{fig:timecut}%
\includegraphics[width=0.5\columnwidth]{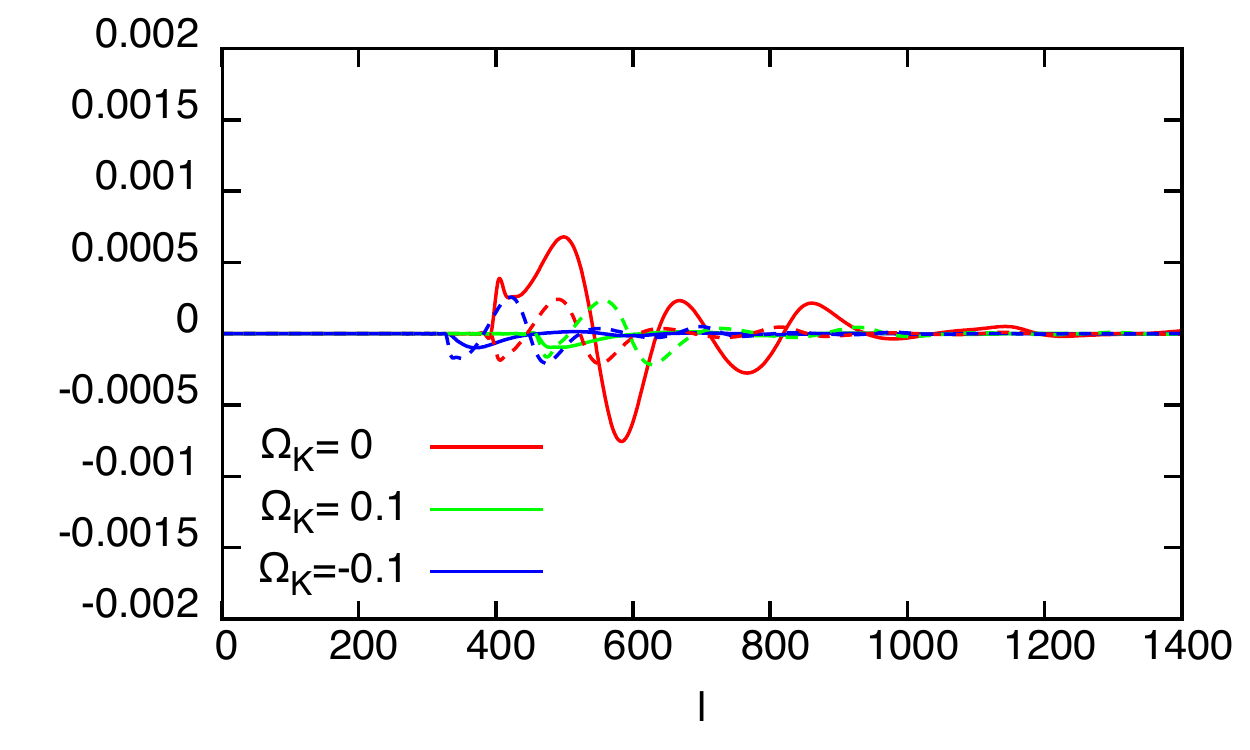}%
\includegraphics[width=0.5\columnwidth]{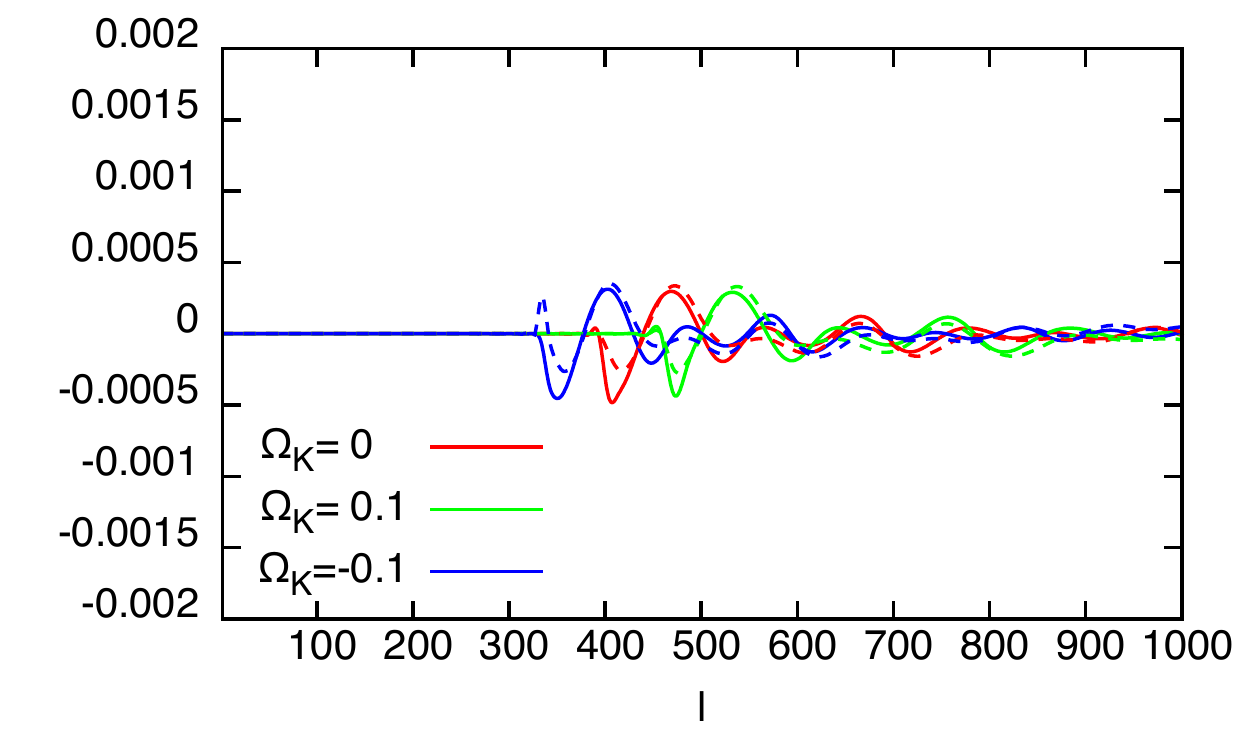}%
\caption{Relative difference between unlensed $C_\ell$'s computed using the time cut approximation, with $(g_c, l_c)=(10^{-3}, 400)$, over $C_\ell$'s without this approximation, for scalar temperature (left, solid), scalar E-type polarisation (left, dotted), tensor E-type polarisation (right, solid) and tensor B-type polarisation (right, dotted), each time for three values of spatial curvature. All calculations have been performed using the reference settings of the file {\tt cl\_ref.pre} for all precision parameters but $(g_c, l_c)$. The vertical scale is the same as in the previous figure.}%
\end{figure}

In Figure \ref{fig:timecut}, we show that the impact of the time cut approximation on scalar and tensor $C_\ell$'s is well below the 0.1\% level for default settings. The advantage of this approximation is that the time spent by the code in calculating the transfer functions for $T_1^{(0)}$, $T_2^{(0)}$, $E^{(m)}$, $B^{(2)}$ is significantly smaller than the time spent in calculating $T_0^{(0)}$ and $T_2^{(2)}$. In the new approach of \CLASS{} v2.0 for source functions, described in section~\ref{sec:sources}, we have increased the number of temperature source term from one to three for scalar modes; thanks to the time cut approximation, this extension is done at reduced cost and does not compromise the speed of the code.

\section{Multipole cut approximation \label{sec:multipole_cut}}

The execution time of Boltzmann codes depends crucially on the number of time integrals that must be performed in order to compute each transfer function ${\Delta_l^X}^{(m)}(q)$. Usually, a comparable amount of time is spent in integrating the system of cosmological perturbations, and in evaluating the transfer functions\footnote{in \CLASS{}, these two tasks are distributed respectively to the {\tt perturbation} and {\tt transfer} modules. Depending on the input cosmological model and on the requested output, most of the time can be spent in one or the other module. In non-flat space, for large enough $|\Omega_k|$, most of the time is actually spent in computing hyperspherical Bessel functions, but the next longest tasks remain the two previous ones.}. Hence, limiting the number of discrete values $l_i$ and $q_j$ for which transfer functions are calculated has a great potential in speeding up the codes.

The CMB transfer functions ${\Delta_l^X}^{(m)}(q)$  (with $X \in \{T_0, T_1, T_2, E, B\}$) peak close to the (generalised) wavenumber $q$ corresponding to Fourier modes seen under an angle $\pi/l$ when they propagate orthogonally to the line-of-sight on the last scattering surface. For a given $l$, these modes are given by $q = q(l) \equiv l/r_A^\mathrm{rec}$. For $q\ll q(l)$, the $\Delta$'s vanish exponentially, because larger wavelengths cannot project under the angle $\pi/l$. For $q\gg q(l)$, the transfer function exhibits damped oscillations, because smaller wavelength can be seen under the same angle if they propagate with an appropriate angle with respect to the line-of-sight. Depending on the type $X$ and mode $(m)$, the transfer function decreases faster or slower with $q$, because some transfer functions can also receive a physical contribution from much smaller wavelengths than $q(l)$ at smaller redshifts, due to the ISW effect or to reionisation. 

For the reasons mentioned in the previous paragraph, the ${\Delta_l^X}^{(m)}(q)$ should not be computed over a rectangular shape in $(q,l)$ space, but around an oblique band encompassing the line $l = q r_A^\mathrm{rec}$. In \CLASS{} v2.0 the most exterior loop is over $q$, so we must express this condition in terms of minimal and maximal values of $l$ for which ${\Delta_l^X}^{(m)}(q)$ should be computed, given the wavenumber $q$, the type $X$ and the mode $m$. 

The minimal $l$ values are easy to find, and are defined in the same way in all versions of \CLASS{}. In section~\ref{sec:time_cut},
we have seen that for each integral over time, the lower boundary is fixed by a threshold value of the Thomson scattering rate, and the upper boundary by a threshold value of the Bessel function (unless the time cut approximation imposes a stronger condition). For $l$ slightly bigger than $q r_A^\mathrm{rec}$, these two boundaries cross each other, because the support of the Bessel function and of the source function do not overlap. In this situation, the code does not perform any integration over time, and simply assigns zero to the transfer function.

\begin{table}\label{tab:lcut}%
\begin{center}
\begin{tabular}{|l|ccccccc|}
\hline
$X^{(m)}$ & $T_0^{(0)}$ & $T_1^{(0)}$ & $T_2^{(0)}$ & $E^{(0)}$ & $T_2^{(2)}$ & $E^{(2)}$ & $B^{(2)}$ \\ 
$(\Delta q)_X^{(m)}$ & 0.15 & 0.04 & 0.15 & 0.11 & 0.20 & 0.25 & 0.10\\
\hline
\end{tabular}
\caption{Default setting for the multiple cut approximation, for scalar and vector modes.}
\end{center}
\end{table}
\begin{figure}\label{fig:lcut}%
\includegraphics[width=0.5\columnwidth]{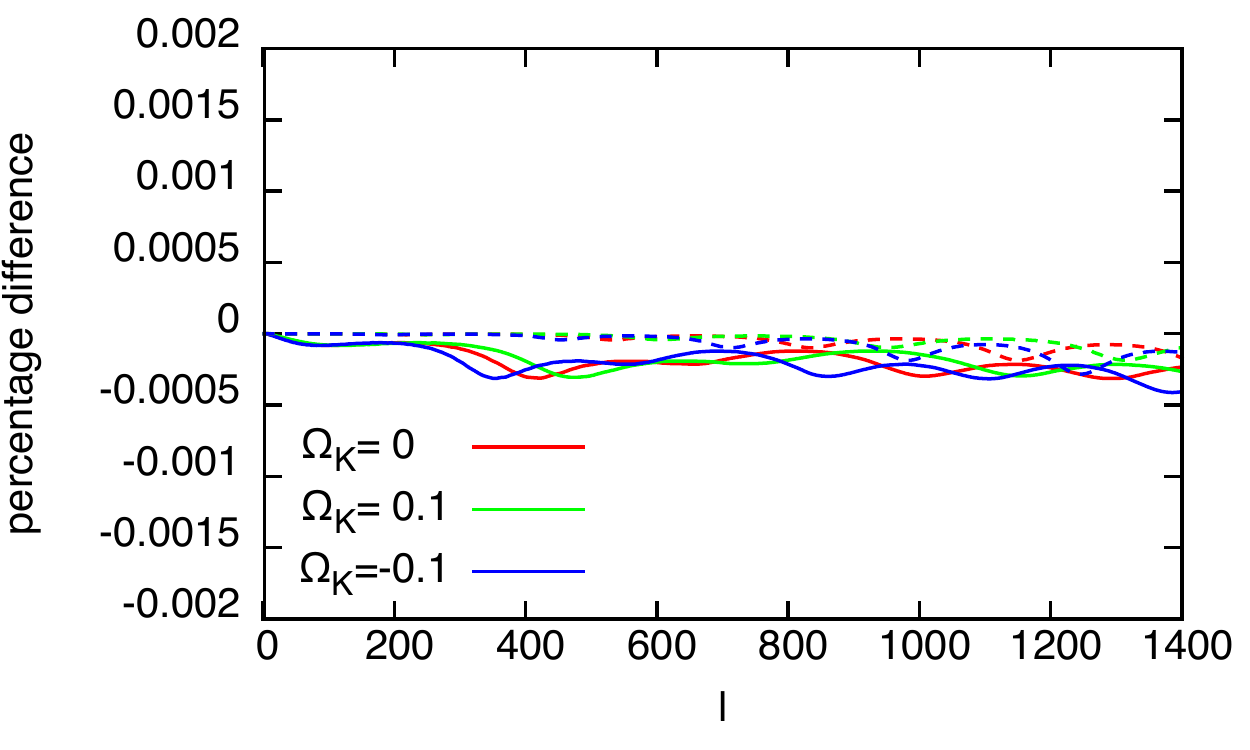}%
\includegraphics[width=0.5\columnwidth]{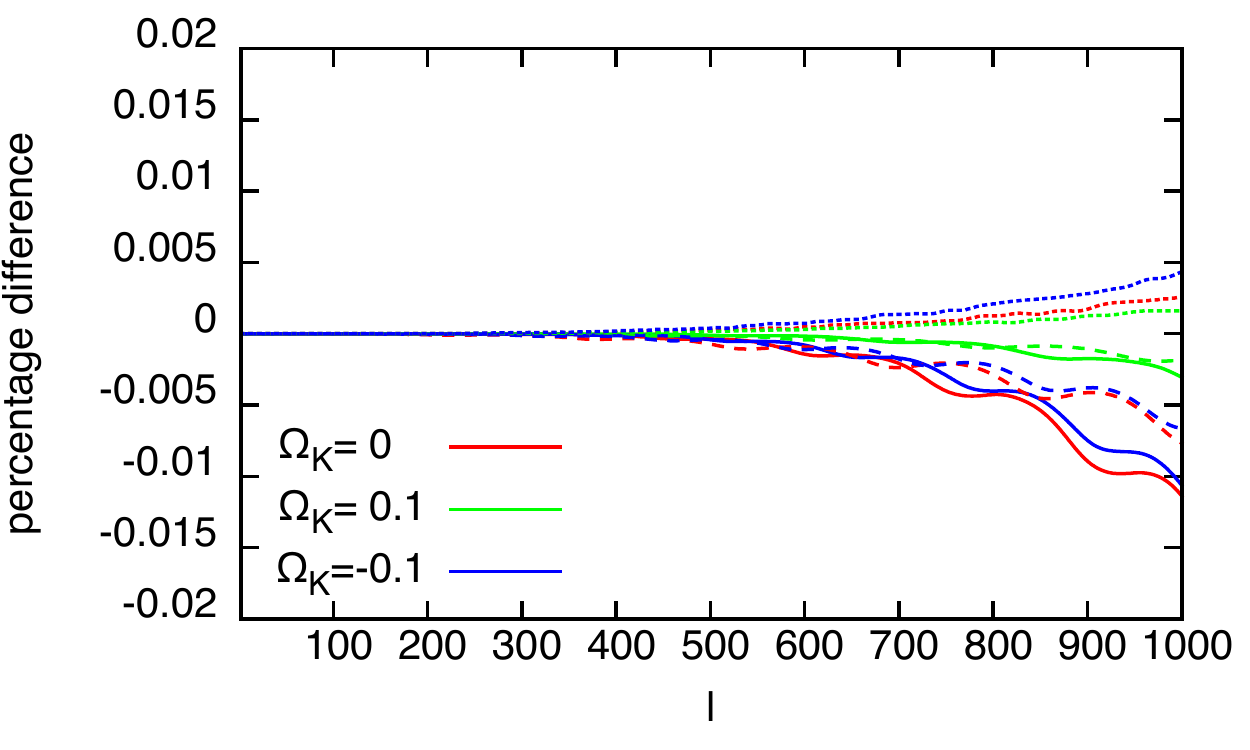}%
\caption{Relative difference between unlensed $C_\ell$'s computed using the multipole cut approximation, with the settings of Table~1, over $C_\ell$'s without this approximation, for scalar temperature (left, solid), scalar E-type polarisation (left, dashed), tensor temperature (right, solid), tensor E-type polarisation (right, dashed) and tensor B-type polarisation (right, dotted), each time for three values of spatial curvature. All calculations have been performed using the reference settings of the file {\tt cl\_ref.pre} for all precision parameters but the $(\Delta q)_X^{(m)}$'s. For scalars, the vertical scale is the same as in previous figures, while for tensors (requiring less precision) the scale is multiplied by ten.}%
\end{figure}
The scheme for the maximal $l$ has been simplified in \CLASS{} v2.0 with respect to versions 1.x. We simply assume a linear boundary for the region in $(q,l)$ space where ${\Delta_l^X}^{(m)}(q)$ is not negligible. In other words, we set the transfer function to zero whenever  the condition $l < \left(q-(\Delta q)_X^{(m)}\right) r_A^\mathrm{rec}$ is fulfilled, where the $(\Delta q)_X^{(m)}$'s are precision parameters tuned to achieve a given precision\footnote{in the code, the parameters $(\Delta q)_X^{(m)}$ are called, e.g., {\tt transfer\_neglect\_delta\_k\_S\_t0} (for scalar modes and type $T_0$).}. Table~\ref{tab:lcut} shows the default settings of \CLASS{} v2.0, which are sufficient for fitting Planck data. Indeed, Figure~\ref{fig:lcut} shows that these settings introduce an error well below the 0.1\% level for all scalar ${C_\ell^{XX}}^{(0)}$'s. Primordial tensor spectra are not yet observed and require less precision, especially on small angular scales. For all ${C_\ell^{XX}}^{(2)}$'s, Figure~\ref{fig:lcut} shows that the error is below 0.1\% for $l<500$, and increases to the level of 1\% for $l\sim 1000$. 

The user is free to avoid ever using the multiple cut approximation by setting all precision parameters $(\Delta q)_X^{(m)}$ to very large values, as done in the reference settings of the file {\tt cl\_ref.pre}.

\section{Limber approximation \label{sec:limber}}

In both the flat and curved space, we use the Limber approximation for speeding up the calculation of the lensing potential transfer function $\Delta_l^\phi(q, \tau_0)$ for large $l$ (by default, $l>10$). This approximation is very useful for lensing, because of the slow variation of the source function with respect to (hyper)spherical Bessel functions. It is never used for CMB transfer functions, due to the quickly oscillating behaviour of the underlying source functions. 

In flat space, the Limber approximation can be derived by Taylor expansion of the function that we are convolving with the spherical Bessel function:
\begin{align}
\int_0^\infty dx j_l(x) f(x) &\simeq \int_0^\infty dx j_l(x) \left( f(x_0) + f'(x_0) (x-x_0) + \cdots \right) \\
&= f(x_0) \int_0^\infty dx j_l(x) + f'(x_0) \int_0^\infty dx j_l(x) (x-x_0) + \cdots
\end{align}
The second integral vanishes if we take
\begin{align}
x_0 &= (l+1)\frac{\Gamma\left( \frac{l+2}{2} \right)^2}{\Gamma\left( \frac{l+1}{2} \right) \Gamma\left( \frac{l+3}{2} \right)} \\
    &\simeq \frac{1}{2} + l + \frac{1}{8l} - \frac{1}{16 l^2} + \cdots
\end{align}
where the last expansion is valid for large $l$. The Limber approximation now becomes
\begin{align}
\int_0^\infty dx j_l(x) f(x) &\simeq f(x_0) \frac{\sqrt{\pi}}{2} \frac{\Gamma\left( \frac{l+1}{2} \right)}{\Gamma\left( \frac{l+2}{2} \right)} \equiv f(x_0) I_l^\text{flat} \\
&\simeq f(x_0) \sqrt{\frac{\pi}{2l}} \left\{ 1 - \frac{1}{4l}  + \frac{1}{32l^2} + \cdots \right\}
\end{align}
We should note that one can also derive a flat space Limber approximation based on a Taylor expansion of $f(x)/\sqrt{x}$. The advantage is that the spherical Bessel function becomes an ordinary Bessel function of order $l+1/2$ and the Limber approximation can be derived to all orders using the Laplace transform~\cite{LoVerde:2008re}. This leads to the Limber approximation
\begin{align}
\int_0^\infty dx j_l(x) f(x) &\simeq \frac{f(\tilde{x}_0)}{\sqrt{\tilde{x}_0}} \sqrt{\frac{\pi}{2}} = f(l+1/2) \sqrt{\frac{\pi}{2l+1}}.
\end{align}
By comparing the two formulae, it is clear that they are equivalent for large $l$. However, the first approach is more easily generalised to curved space. First note that the classical turning point is a good approximation to $x_0$ as is the case in flat space where $x_\text{tp}=\sqrt{l(l+1)} \simeq l + \frac{1}{2} - \frac{1}{8l} + \cdots$. We found numerically that the integrals over the hyperspherical Bessel functions could be expressed in terms of the flat integrals $I_l^\text{flat}$ to a good approximation as
\begin{align}
\int_0^\infty d\chi \Phi_l^\nu(\chi) &\simeq \left[1 - \kk \frac{l^2}{\nu^2} \right]^{-1/4} \frac{1}{\nu} I_l^\text{flat},
\end{align}
where $\infty$ in the integrals should be understood as the equivalent point $\pi/2$ for $\kk=1$. The Limber approximation in curved space now becomes
\begin{equation}
\int_0^\infty d\chi \Phi_l^\nu (\chi) f(x) \simeq f(\chi_0) \frac{1}{\nu}  \left(1 - \hat{K} \frac{l^2}{\nu^2} \right)^{-1/4} I_l^\text{flat} 
\end{equation}
Like in flat space, we choose $\chi_0= \mathrm{Arcsin}_{\hat{K}} \left(\frac{l + \frac{1}{2}}{\nu}\right)$, very close to the turning point $\chi_\mathrm{tp}$ defined in section \ref{sec:ahbf}. Then we can derive:
\begin{eqnarray}
\Delta_l(q) &=& \int d \tau S(k,\tau) \Phi_l^{\nu=q/\sqrt{|K|}}\left(\chi=\sqrt{|K|}(\tau_0-\tau) \right)\\
&=& \int d \chi \frac{1}{\sqrt{|K|}} S\left(k,\tau=\tau_0-\frac{\chi}{\sqrt{|K|}}\right) \Phi_l^{\nu=q/\sqrt{|K|}}(\chi) \\
&\simeq & I_l^\text{flat} \left(1 - \hat{K} \frac{l^2}{\nu^2} \right)^{-1/4} \frac{1}{q} S\left(k,\tau=\tau_0-\frac{\chi_0}{\sqrt{|K|}}\right) 
\end{eqnarray}
This approximation is nearly the same as the one implemented in {\sc camb}, up to tiny corrections (terms $+1$ or $+\frac{1}{2}$).
Note that in the code, the function that we want to interpolate with respect to time with is not $S$ but the product $(\tau_0-\tau) S$: this product is better behaved in $(\tau_0-\tau) \longrightarrow 0$. Like in flat space, we push the calculation one step further, because we want to interpolate $(\tau_0-\tau) S$ instead of $S$:
\begin{eqnarray}
\Delta_l(q)
&\simeq&
I_l^\text{flat} \left(1 - \hat{K} \frac{l^2}{\nu^2} \right)^{-1/4} \frac{1}{q} \frac{\sqrt{|K|}}{\chi_0} \left[ (\tau_0-\tau) S(k,\tau) \right]_{\tau =\tau_0-\frac{\chi_0}{\sqrt{|K|}}} \\
&=&
I_l^\text{flat} \left(1 - \hat{K} \frac{l^2}{\nu^2} \right)^{-1/4} \frac{1}{\nu \chi_0} \left[ (\tau_0-\tau) S(k,\tau) \right]_{\tau =\tau_0-\frac{\chi_0}{\sqrt{|K|}}} ~.
\end{eqnarray}
This version of the Limber approximation in non-flat space has been implemented in \CLASS{} v2.3.3.

\bibliographystyle{utcaps}

\bibliography{superhyperpaperreferences}

\end{document}